%

\def\W#1{#1_{\kern-3pt\lower6.6truept\hbox to 1.1truemm
{$\widetilde{}$\hfill}}\kern2pt\,}
\font\titolo=cmr12\font\dodiciit=cmti12
\def\DD{{\cal D}}\def\LL{{\cal L}}
\def\FF{{\cal F}}\let\st=\scriptstyle
\def\lis#1{{\overline #1}}\def\fra#1#2{{#1\over#2}}
\def\rif#1#2#3#4{{}}\def\raf#1#2#3#4{\noindent[#4] #3\par\*}
\def\*{\vskip3mm}


\let\a=\alpha \let\b=\beta \let\g=\gamma \let\d=\delta
\let\e=\varepsilon \let\z=\zeta \let\h=\eta
\let\th=\vartheta\let\k=\kappa \let\l=\lambda \let\m=\mu \let\n=\nu
\let\x=\xi \let\p=\pi \let\r=\rho \let\s=\sigma \let\t=\tau
 \let\ch=\chi  \let\o=\omega
  \let\D=\Delta 
 \let\P=\Pi \let\Si=\Sigma \let\F=\Phi
 \let\O=\Omega 

\def\equ{{}}\def\fra#1#2{{#1\over#2}}\def\V#1{{\,\underline#1\,}}

\let\0=\noindent
\hsize=14.5cm\vsize=22.cm
\newdimen\xshift \newdimen\xwidth \newdimen\yshift
\def\ins#1#2#3{\vbox to0pt{\kern-#2 \hbox{\kern#1
#3}\vss}\nointerlineskip} \def\eqfig#1#2#3#4#5{ \par\xwidth=#1
\xshift=\hsize \advance\xshift by-\xwidth \divide\xshift by 2
\yshift=#2 \divide\yshift by 2 \line{\hglue\xshift \vbox to #2{\vfil #3
\includegraphics{#4.ps} }\hfill\raise\yshift\hbox{#5}}} 
\def\8{\write13}
\def\figini#1{\catcode`\%=12\catcode`\{=12\catcode`\}=12
\catcode`\<=1\catcode`\>=2\openout13=#1.ps}
\def\figfin{\closeout13\catcode`\%=14\catcode`\{=1
\catcode`\}=2\catcode`\<=12\catcode`\>=12}
\headline{Physica D, {\bf 105}, 163--184, 1997.\hfill}
\vskip2.5cm

\centerline{\titolo Dynamical ensembles equivalence}
\centerline{\titolo in fluid mechanics\footnote{*}{\rm{\it
mp$\_$arc@math.utexas.edu}, \#96-182, and
{\it chao-dyn@xyz.lanl.gov},\#9605006}}
\vskip1.cm
\centerline{\dodiciit G. Gallavotti}
\centerline{Fisica, Universit\`a
di Roma La Sapienza, P.le Moro 2, 00185, Roma, Italia.}
 
\vskip1cm
 
\0{\it Abstract: An attempt to put togheter various theoretical,
mathematical, or experimental results recently developed in apparently
unrelated subjects.  Namely Ruelle's approach to turbulence, [1], the
body of Nos\'e--Hoover type of molecular dynamics experiments,
[2],[3],[4], mathematical results on Lyapunov exponents (the pairing
rule, [5],[6]) and experimental results on them ([7],[8])), theoretical
as well as mathematical results on fluctuations ([9] (multifractality),
and [10] (chaotic hypothesis)).  The key idea that we
try to clarify is that of "dynamical ensembles", as a generalization of
the classical "equilibrium ensembles", arguing that they should be
identified with the SRB distributions and that they share several
properties with the classical ensembles.  Most of the results invoked
here did not deal directly with the Navier Stokes equations and yet they
seem to have a lot to do with them (as we shall argue): here the
discussion will focus on the Navier Stokes and dissipative Euler
equations with the aim of proposing several experiments apt to test the
equivalence of dynamical ensembles and the chaotic hypothesis.  The
ideas developed, to a great extent, from the efforts put in interpreting
the experimental results in [4]. An Erratum has been added before
the references.}
 
\*
\0{\sl PACS: 47.52.+j, 05.45.+b, 05.70.Ln, 47.70.-n}
 
\0{\sl Keywords: Navier Stokes, Chaotic hypothesis, Caos, Nonequilibrium
ensembles}
 
\vskip1truecm
\0{\it\S1 Reversible dissipation in Euler and Navier Stokes equations.}
\*
In [11] one finds an analysis leading to a conjecture on the
equivalence between the irreversible NS equation and a reversible
equation that was called GNS ("gaussian NS equation").
 
The ideas discussed in [11] are, however, much more general
and it is worth pointing out some other applications, as well as
several possible tests that seem under reach of present day
experimental (numerical {\it and} "real") techniques.
 
The paper freely uses mixed results available in the literature.
Therefore the reader who, after a first reading, has still some interest
in the matter may find it difficult to distinguish between my
conjectures or theorems, and other's conjectures, results and/or
theorems, and he may want to avoid a second reading and checking of the
references to find out. Therefore I have added a concluding section where
this is spelled out again in a concise form. The concluding section can
also be read, by a reader familiar with the subject,
right after seeing the equations \equ(1.1): this will provide an informal
but detailed overview of what is discussed in the paper, and of its
relations with other works.
 
I shall focus on fluid mechanics problems considering
a fluid that:
 
\0(1) is enclosed in a periodic box $\O$ with side $L$, possibly
with a few disks ("obstacles") removed so that no infinite straight
path can be found in $\O$ that avoids the obstacles,
 
\0(2) is incompressible with density $\r$.
 
I shall consider four distinct evolution equations for this fluid, all
of dissipative nature.
\vfill\eject
 
$$\eqalign{
\dot{\V u}+\W u\cdot\W \partial \,\V u=-\fra1\r\V\partial p+\V g+\n\D\V u,
\qquad\V\partial\cdot\V u=0&\qquad {\rm NS}\cr
\dot{\V u}+\W u\cdot\W \partial \,\V u=-\fra1\r\V\partial p+\V g+\b\D\V u,
\qquad\V\partial\cdot\V u=0&\qquad {\rm GNS}\cr
\dot{\V u}+\W u\cdot\W \partial \,\V u=-\fra1\r\V\partial p+\V g-\ch\V u,
\qquad\V\partial\cdot\V u=0&\qquad {\rm ED}\cr
\dot{\V u}+\W u\cdot\W \partial \,\V u=-\fra1\r\V\partial p+\V g-\a\V u,
\qquad\V\partial\cdot\V u=0&\qquad {\rm GED}\cr
}\eqno(1.1)$$
 
{\it In the case $\O$ contains obstacles a "no friction" boundary
condition will be imposed on $\partial \O$,} {\it i.e. } $\V u\cdot\V
n=0$ if $\V n$ is the normal to $\partial\O$.  The first equation is the
well known Navier Stokes equation with $\n$ being the {\it viscosity}.
 
The second equation, introduced in [11] and called GNS, has a
multiplier $\b$ defined so that the total vorticity $\h L^3=\r\int \V\o^2\,dx$,
with $\V\o=\partial\wedge\V u$ being the {\it vorticity}, is a constant of
motion; this means that:
 
$$\b(\V u)=\fra{\int\big(\V\partial\wedge
\V g\cdot\V\o+ \V\o\cdot \,(\W \o\cdot\W \partial\V
u)\big)\,d\V x}{\int({{\V\partial}}\wedge\V\o)^2\,d\V x}\eqno(1.2)$$
 
The third equation will be called the {\it Euler dissipative} equation,
ED: it represents a non viscous ideal fluid moving in a "sticky
background": $\ch$ is a "sticky" viscosity.  The model is not, as far
as I know, a good model for any physical situation ({\it i.e.} for
$3D$--fluids), but it is interesting to consider it for comparison
purposes.  In $2D$--fluids the sticky viscosity has interest in its own
as it appears in geophysical models where the coefficient $\ch$ is
known as the {\it Eckman viscosity}.  \footnote{1}{ I am indebted to
the referee for this comment and for an appropriate reference, [13].}
  
The fourth equation will be called GED equation, {\it gaussian
dissipative Euler} equation and here $\a$ is a multiplier defined so
that the total (kinetic) energy $\e L^3=\fra\r2\int\V u^2\,dx$ is a
constant of motion {\it in spite} of the action of the force $\V g$;
this means that $\a$ is given by:
 
$$\a(\V u)=\fra{\int \V g\cdot\V u}{\int \V u^2}\eqno(1.3)$$
{\it A similar equation, with the constraint that the energy contained in
each ``momentum shell'' be a constant}, was considered in [12].
which is the first paper in which the idea of a reversible Navier
Stokes equation is advanced and studied.  The energy content of each
``momentum shell'' was fixed to be the value predicted by Kolmogorov
theory, [14].
 
Note that both the GED and the GNS equations have a symmetry in $\V
u$, so that they are {\it reversible} in the sense that, if $V_t$ is the
flow describing the equation solution (so that $t\to V_t \V u=\V u(t)$
is the solution with initial data $\V u$), then the transformation
$i:\V u\to -\V u$ {\it anti-commutes} with the time evolution $V_t$:
$$i\,V_t\,=\,V_{-t}\,i\eqno(1.4)$$
We shall avoid (as it is, unfortunately, always the case in the current
literature) considering the problem of proving the global existence and 
regularity of solutions to the equations \equ(1.1) (the problem is in
fact {\it open}, see [15]) and we shall consider the {\it
truncated equations} with momentum cut off $K$.
 
The truncation will be performed on a suitable orthonormal basis for
the {\it divergenceless} fields in $\O$: given the boundary conditions
we consider it natural to use the basis generated by the {\it minimax}
principle applied to the Dirichlet quadratic form $\int_\O (\W\partial\,\V
u)^2d\V x$ defined on the space of the $C^\infty(\O)$ divergenceless
fields $\V u$ with $\V u\cdot\V n=0$ on $\partial\O$.  The basis fields $\V
u_j$ will verify: $\D\V u_j=-E_j\V u_j+\V\partial\m_j$, for a suitable
multiplier $\m_j$, with $\V u_j,\m_j\in C^\infty$ and $E_j$ are
eigenvalues).
 
For instance in the case of {\it no obstacles} let the $\V u=\sum_{\V
k\ne\V0}\V\g_{\V k} e^{i\V k\cdot\V x}$ be the velocity field
represented in Fourier series with $\V \g_{\V k}=\lis{\V \g_{-\V k}}$
and $\V k\cdot\V\g_{\V k}=0$ (incompressibility condition); here $\V k$
has components that are integer multiples of the "lowest momentum"
$k_0=\fra{2\p}L$.  Then consider the equation:
 
$$\dot{\V\g}_{\V k}=-\th({\V k})\V \g_{\V k} -i\sum_{\V k_1+\V k_2=\V k}
(\V\g_{\V k_1}\cdot\V k_2)\, \P_{\V k}\V \g_{\V k_2}+\V g_{\V k}
\eqno(1.5)$$
where the $\V k$'s take only the values $0<|\V k|<K$ for some {\it
momentum cut--off} $K>0$ and $\P_{\V k}$ is the projection on
the plane orthogonal to $\V k$. This is an equation that defines a
"truncation on the momentum sphere with radius $K$ of the equations
\equ(1.1)" if:
 
$$\eqalign{
\th({\V k})=-\n {\V k}^2\qquad & {\rm  NS\ case}\cr
\th({\V k})=-\b {\V k}^2\qquad & {\rm GNS\ case}\cr
\th({\V k})=-\ch \qquad & {\rm ED\ case}\cr
\th({\V k})=-\a \qquad &{\rm GED\ case}\cr}\eqno(1.6)$$
For simplicity we may suppose, in this no obstacles cases, that the mode
$\V k=\V 0$ is {\it absent}, {\it i.e. } $\V\g_{\V0}=\V0$: this can be done if,
as we suppose, the external force $\V g$ does not have a zero mode
component ({\it i.e. } if it has zero average).
 
In order that the resulting cut--off equations be physically acceptable,
and supposing that $\V g_{\V k}\ne\V0$ only for $|{\V k}|\sim k_0$, one shall
have to fix $K$ large. For instance in the NS case it should be
much larger than the {\it Kolmogorov scale} $K=(\h\n^{-2})^{1/4}$,
where $\n\h$ is the average dissipation rate of the solutions to \equ(1.5)
without cut--off.  The scale $K$ is determined, on the basis of
heuristic dimensional considerations and of the dissipation rate
$\n$--independence (as $\n\to0$: [16] p.  306), by setting
$\n\h\sim L^{1/2}g^{3/2}$): see [14].
 
We shall use the same cut off for the other equations with ${\V k}$
replaced by the basis label $j$ and $|{\V k}|$ replaced by $\sqrt{E_j}$,
which is certainly a natural choice for the GNS equation.
 
For the ED and GED equations the choice of $K$ should be made by
developing a theory analogous to Kolmogorov's theory.  We only attempt
a preliminary analysis in \S6 as the latter equations are used here
only for the purpose of illustrating some interesting mechanisms and
theories.  Below we always refer to the truncated equations, unless
otherwise stated.
 
It is easy, in the no obstacles cases, to express the coefficients
$\a,\b$ for the cut off equations:
 
$$\eqalignno{
&\a=\fra{\sum_{0<{\V k}|<K}\lis {\V g}_{\V k}\cdot\V\g_{\V k}}{
\sum_{0<{\V k}|<K}\V\g_{\V k}^2}\qquad \b=\b_i+\b_e
            &(1.7)\cr
&\b_e=\fra{\sum_{\V k\ne\V0}{\V k}^2\V g_{\V k}\cdot
\lis{\V \g}_{\V k}}{\sum_{\V k}\V k^4|\V \g_{\V k}|^2}\cr
&\b_i=\fra{-i\sum_{\V k_1+\V k_2+\V k_3=\V0}
\V k_3^2\,(\V\g_{\V k_1}\cdot\V k_2)\,(\V \g_{\V k_2}\cdot\V \g_{\V
k_3})}{\sum_{\V k}\V k^4|\V \g_{\V k}|^2}\cr}$$
where the $\V k$'s take only the values $0<|\V k|<K$ for some {\it
momentum cut--off} $K>0$ and $\P_{\V k}$ is the orthogonal projection on
the plane perpendicular to $\V k$.
 
\*
The cases in which the region $\O$ contains obstacles is very similar
but we cannot write simple expressions for the basis fields and
therefore the equations, although formally very similar to \equ(1.5),
\equ(1.7), cannot be written very explicitly.
 
The solutions of the equations \equ(1.5), or of the corresponding ones
in the obstacles cases, will be denoted $V^{\n,ns}_t\V u,V^{\h,gns}_t\V
u,V^{\ch,ed}_t\V u, V^{\e,ged}_t\V u$ when the initial datum is $\V u$. 
Or in general:

$$V^{\x}_t\V u,\quad\x=(\n,ns),\,(\h,gns),\,(\ch,ed),\,(\e,ged)\eqno(1.8)$$

Keeping the forcing $\V g$ constant we shall admit that for each
equation, {\it i.e. } for each choice of $\x$, there is a unique stationary
distribution $\m_{\x}$ describing the statistics of all initial data
$\V u$ that are randomly chosen with a "Liouville distribution", {\it i.e. }
(in the no obstacle cases, to fix the ideas) with a distribution
$\m_0(d\V \g)$ proportional to the volume measure $\d_\x\, \prod_{|\V
k|<K}d \V\g_{\V k}$, where the delta function is present only in the
case of the reversible equations and fixes the constants of motion to
the value prescribed by the first label in $\x$.
 
This means that given "any observable" $F$ on the phase space $\FF$ (of
the velocity fields with momentum cut--off $K$) it is:

$$\lim_{T\to\infty}\fra1T\int_0^T F(V^{\x}_t \V \g) dt=
\int_\FF F(\V \g') \m_{\x}(d\V \g'){\buildrel def \over =}
\langle\,F\,\rangle_\x\eqno(1.9)$$
for all choices of $\V \g$ except a set of zero Liouville measure.  The
distribution $\m_\x$ will be called the SRB distribution for the
Eq. \equ(1.5),\equ(1.6) (see the "zero-th law" in
[17],[18]).
 
A particular role will be plaid by the averages $\langle\,\e\,\rangle_\x,
\langle\,\h\,\rangle_\x$ as well as by the averages of
$\langle\,\a\,\rangle_\x,\langle\,\b\,\rangle_\x$ and of the {\it entropy production rate}
$\s(\V\g)$, that is {\it defined by the divergence of the r.h.s.  of
the cut off equations}.
 
Consider explicitly only the no obstacles case: if $D_K$ is the number
of modes ${\V k}$ with $0<|{\V k}|<K$ then the number of (independent)
components of $\{\V\g_{\V k}\}$ is $2D_K$ and, see \equ(1.5), setting
$2\lis D_K=\sum_{|\V k|<K}2\V k^2$ (which in the case with obstacles
become $2\lis D_K=\sum_{\sqrt{E_j}<K}\sqrt{E_j})$), one finds that $\s$
is given by:
 
$$\eqalign{
\s=& 2\lis D_K\n\qquad\kern1.5cm\x=(\n,ns)\cr
\s=& 2\lis D_K\b-\lis \b_e-\lis\b_i\qquad \x=(\h,gns)\cr
\s=& 2 D_K\ch\qquad \kern1.5cm\x=(\ch,ed)\cr
\s=& 2D_K\a-\a\kern1.5cm\x=(\e,ged)\cr}\eqno(1.10)$$
where $\lis\b_i,\lis\b_e$ are suitably defined, {\it e.g.} in the no obstacles
cases:
 
$$\lis\b_e =\fra{\sum_{\V k} \V k^2\lis{\V g}_{\V k}\cdot\V\g_{\V k}}
{\sum_{\V k}\V k^4
|\V \g_{{\V k}}|^2}-
2\fra{\big(\sum_{\V k} \V k^2\lis{\V g}_{\V k}\cdot\V\g_{\V k}\big)
\big(\sum_{\V k} \V k^2\V\g_{\V k}^2\big)}{\big(\sum_{\V k}\V k^4
|\V \g_{{\V k}}|^2\big)^2}\eqno(1.11)$$
so that $\s\simeq 2\lis D_K\b$ for $\x=(\h,gns)$ and $\s\simeq2 D_K\a$
for $\x=(\e,ged)$.
 
The {\it equivalence of dynamical ensembles conjecture}, [11],
is the following:\footnote{2}{...  \`e tanto nuova e, nella prima
apprensione, remota dal verisimile, che quando non si avesse modo di
dilucidarla e renderla pi\`u chiara che'l Sole meglio sarebbe il
tacerla che'l pronunziarla; per\`o, gi\`a che me la son lasciata
scappare di bocca..., [19], p.  231.}
\*
 
{\it Conjecture NS: The statistics $\m_{\n,ns},\m_{\h,gns}$ of the NS
equations and of the GNS equations respectively are {\sl equivalent} in
the limit of large Reynolds number {provided} the parameters $\h$ and
$\n$ are so related that $\langle\,\s\,\rangle_{\n,ns}=\langle\,\s\,\rangle_{\h,gns}$.} \*
 
\0Here {\it equivalent} means that the ratios of the averages of the
same observables with respect to the two distributions approaches $1$
as $R\to\infty$.  The Reynolds number is defined here to be
$R={(\n\h)}^{1/3}L^{4/3}\n^{-1}$ because the condition of equivalence
can also be expressed by: $\n\langle\,\V\o^2\,\rangle=\n\h$ and the quantity
$\n\h$ is the quantity that in the usual notatins of Kolmogorov's
theory is called $\e$.  We do not adhere to such notation only to avoid
confusion with the quantity called (naturally) $\e$ ("energy per unit
mass") in the analysis of the ED and GED equations.
 
A corresponding conjecture can be formulated for the ED and GED
equations:
\*
 
{\it Conjecture ED: The statistics $\m_{\ch,ed},\m_{\e,ged}$
of the ED equations and of the GED equations respectively are {\sl
equivalent} in the limit of large Reynolds number {provided}
the parameters $\e$ and $\ch$ are so related that
$\langle\,\s\,\rangle_{\ch,ed}=\langle\,\s\,\rangle_{\e,ged}$.}
\*
 
The above stated conjectures are closely analogous to the familiar
statements on the equivalence of thermodynamic ensembles, with the
thermodynamic limit replaced by the limit $R\to\infty$ of infinite
Reynolds number.  They can be substantially weakened for the purposes
of possible applications.
 
It is well known that the equivalence of the ensembles in equilibrium
statistical mechanics {\it does not extend} to all possible
observables, but it is restricted to the {\it local} ones.  The natural
notion of locality is, in the cases above, locality in momentum space.
 
An observable $O$ will be called "local" if its value on a particular
velocity field depends only on the Fourier components of the field
with wave vector $\V k$ in a range $k_1,k_2$ {\it independent} on the
Reynolds number size.
 
Typical local observables are the energy content of a momentum shell,
and the average (overs space) velocity near a point.  The difference
between the average velocity field near a point $\V x$ and an
infinitesimally close field can be considered a local observable if
"close" means "differing" little only near $\V x$.  On the other hand
the quantity $\s$, total phase space contraction rate, is a non local
observable.
 
Therefore one can expect that, in equivalent distributions, the
Lyapunov exponents of the two models coincide, at least if one looks at
the ones in a fixed range of values, away from the extreme values.  The
fluctuations of $\s$ may be quite different (although the average
values of this quantity will still be the same, tautologically, if the
conjecture holds).
 
The idea of non equilibrium ensembles and their possible equivalence is
not really new: the recent literature contains many hints in such
direction. The clearest is perhaps [12]. See also the first
of [20] (\S4) and [21].
 
On heuristic grounds, the conjectures would be justified if one did
accept that the entropy creation rate reaches its average on a time
scale that is fast compared to the hydrodynamical scales.  The
coefficients $\a\simeq(2D_K)^{-1}\s$, and $\b\simeq(2\lis D_K)^{-1}\s$,
see \equ(1.10), would be confused with their time averages
$\langle\,\a\,\rangle_{\e,gne}$ or $\langle\,\b\,\rangle_{\h,gns}$ and {\it identified with
the viscosity} constants $\n$ or $\ch$.
 
In this way the GNS and the NS equations would be equivalently good:
both being the macroscopic manifestation of two {\it equivalent
microscopic dissipation mechanisms}: one explicitly specified by the
Gaussian constraint of constant dissipation and the other with
dissipation {\it unspecified {\it a priori}} but phenomenologically modeled by a
constant viscosity.  Likewise one can view the GED and the ED equations
as macroscopically equivalent: one with constant energy and the other
with constant sticky viscosity $\ch$.
 
The interest of the above conjectures is that the same physical system
in which irreversible dissipation occurs (the NS or ED equations)
can be described equivalently by a reversibly dissipative system
(the GNS or GED equations).
 
For instance one can investigate the implications of the fact that for
reversible systems a general principle, {\it the chaotic hypothesis},
can be reasonably assumed to hold and to imply consequences that seem to
be non trivial, see [10], [18], [22], [23], [24], about fluctuations and
Lyapunov spectrum.
 
The next section is devoted to a quick discussion of some of the
established consequences of the principle and \S3\%\S6 will deal with
{\it heuristic} ideas and with describing a possible {\it scenario} for
the phenomenology of the equations \equ(1.1).  The scenario will be
developed {\it without any pretension of rigor} and it will present
what will appear as the {\it simplest} among many other possibilities. 
It leads (implicitly) to several possible experimental tests of the
chaotic hypothesis and of the other ideas involved in its development:
the tests can also be viewed, independently, just as interesting
experiments proposals.
 
In the following we shall always consider the NS equations and the GNS
equations with parameters fixed so that $\m_{\n,ns}$ and $\m_{\h,gns}$
are equivalent by the conjecture NS, and likewise we shall always
consider the ED and GED equations with parameters fixed so that
$\m_{\ch,ed}$ and $\m_{\e,ged}$ are equivalent by the conjecture ED.
\*
\vskip3mm 
{\it\S2 The fluctuation theorems.}
\*
 
In reference [10],[18] the {\it chaotic
hypothesis} was presented as a reformulation of an older principle due
to Ruelle, [1].  It gave us the possibility of some
quantitative parameterless "predictions", in various cases, see also
[22], [23], [24].  The hypothesis is: \*
 
\0{\sl Chaotic hypothesis:} {\it A chaotic many particle system or fluid
in a stationary state can be regarded, for the purpose of computing
macroscopic properties, as a smooth dynamical system with a transitive
Axiom A global attractor.  In reversible systems it can be regarded,
for the same purposes, as a smooth transitive Anosov system.} \*
 
The main result of [10] is the {\it fluctuation theorem} that gives a
property of the variable $p=p(\V\g)$ defined in terms of the
contraction rate $\s_0$ of the attractor surface elements by:

$$\fra1\t\int_{-\t/2}^{\t/2}\s_0(V_t\V\g)\,dt=\langle\,\s_0\,\rangle_+
p\eqno(2.1)$$
which can be regarded as a random variable with the distribution
$\p_\t(p)dp$ that it inherits from $\m_{\h,gns}$ or $\m_{\e,ged}$; here
$\langle\,\s_0\,\rangle_+$ is the average of $\s_0$ with respect to the
distribution $\m_{\h,gns}$ or $\m_{\e,ged}$.
 
Note that $\s_0$ should {\it not be confused} with $\s$, \equ(1.10):
thinking of the attractor as a smooth surface $\s_0$ is the contraction
rate of its surface elements, {\it which is different} from the contraction
rate $\s$ of the phase space volume elements, see \S4\%\S6.
 
If the conjectures of \S1 are accepted $\langle\,\s_0\,\rangle_+$ is also the
$\m_{\n,ns}$ or $\m_{\ch,ed}$ average of $\s_0$ or at least tends to
it as $R\to\infty$.
 
If $\langle\,\s_0\,\rangle_+>0$, see [25] for a discussion of the
conditions for this inequality ("Ruelle's H-theorem"), and if
$\z(p)=\lim_{\t\to\infty}\fra1\t \log \p_\t(p)$ then the {\it fluctuation
theorem} of [10] gives the following {\it large deviation}
relation, see also [22], [26], for the equations
GNS and GED:
 
$$\fra{\z(p)-\z(-p)}{\langle\,\s_0\,\rangle_+p}=1,
\qquad{\rm for\ all\ } p\eqno(2.2)$$
which, in the case of nonequilibrium statistical mechanics, has been
interpreted as an {\it extension of the fluctuation dissipation
theorem} to large forcing fields, [24].  Here "for all" $p$
means for all possible values of $p$ (which is in general a bounded
quantity).
 
The fluctuation theorem \equ(2.2) says that the distribution of $p$ is
{\it multifractal}, not surprisingly since $\z(p)$ can be regarded as a
kind of {\it generalized sum of Lyapunov exponents}
in the sense of [9], [8], and the {\it odd part}
of $\z(p)$ is linear.
 
A more general fluctuation theorem concerns the {\it joint}
distribution of the variable $p$ and of any other variable $q=q(\V\g)$
that is similarly defined in terms of an observable $Q$ which is {\it
odd} under the time reversal operation that is defined on the
attractor, {\it i.e. }:
 
$$\fra1\tau\int_{-\t/2}^{\t/2} dt \,Q(V^\x\V \g)=\langle\,Q\,\rangle_+
\,q\eqno(2.3)$$
If $\p_\t(p,q)$ denotes the joint probability density of the
observables $p,q$ and if $\z(p,q)=\lim_{\t\to\infty}$ $\fra1\t\log \p_\t(p,q)$
then it follows from the chaotic hypothesis that the distributions
for $p,q$, with respect to the statistics $\m_\x$, $\x=(\b,gns)$ or
$\x=(\a,ged)$, verify:

$$\fra{\z(p,q)-\z(-p,-q)}{\langle\,\s_0\,\rangle_+p}=1, \qquad{\rm for\ all\ } p,q
\eqno(2.4)$$
which, in the case of nonequilibrium statistical mechanics, has been
interpreted as an {\it extension of the Onsager's reciprocity}
to large forcing fields, [24].
 
In the case of the equations \equ(1.1), second and fourth, the above
relations are applicable when the motions are chaotic, {\it i.e. } have at
least one positive Lyapunov exponent (which should happen as soon as
$R$ is large enough, excepted possibly very special cases, see
[27]); and it gives an interesting parameterless prediction
{\it if} the contraction rate $\s_0$ can be related to the contraction
rate $\s$ of the GNS equations.
 
If the conjecture in \S1 held the sense of complete asymptotic
equivalence between the ``ensembles'' $\m_{\h,gns}$ and $\m_{\n,ns}$
(or $\m_{\e,ged}$ and $\m_{\ch,ed}$) then \equ(2.2) {\it could also
hold for other models} of the viscous stationary states, like the one
given by the {\it classical NS equation in particular}: note the
quantities $\sigma$ in \equ(1.10) are {\it still} fluctuating variables
even when the evolution considered is given by the NS or ED equation.

This would mean that fixing $\h$ or $\n$ and looking at the
fluctuations of $\s$ could be analogous to fixing the density or the
chemical potential ({\it i.e.} considering the canonical or the grand
canonical ensembles) and looking at the fluctuations of the energy
(which, in the thermodynamic limit, are the same in the two ensembles).

It turns out, from very recent numerical and theoretical results of
F.  Bonetto (private communication) {\it in related but purely mechanical
problems}, that if $\n$ is fixed (in the NS equations) the fluctuations
of $\s$ might be strongly affected and very different from the ones of
the same $\s$ in the case in which (in the GNS equations) $\h$ is fixed
at the right value (as demanded by the conjecture).

Hence one has to be very cautious about extending the "equivalence" to
such relation {\it without} further arguments to exclude that fixing
$\h$ or $\n$ and looking at the fluctuations of $\s$ may be more like
fixing chemical potential and density (in the grand canonical and in
the canonical ensembles) and looking at the density fluctuations in a
box almost as large as the available volume.
  
In the latter case the validity of \equ(2.2) and \equ(2.4) could
only become a way of distinguishing the two equivalent distributions
$\m_{\h,gns},\m_{\n,ns}$.  Just in the same way as in equilibrium the
energy fluctuations distinguish the microcanonical ensemble from the
equivalent canonical ensemble (or density fluctuations distinguish the
canonical and the corresponding grand canonical ensembles).  The lack
of equivalence only affects the fluctuations of "global" quantities
like $p,q$: but this makes necessary to be more precise and to specify
the class of observables for which the equivalence can be conjectured
as done above in the remarks following the conjectures.
 
In real systems a test would require experimental ability of measuring
total vorticity (or total energy) fluctuations.  This is apparently
{\it very difficult} and in any event one cannot hope to measure the
total vorticity (or energy) but only the amount of vorticity (or
energy) contained between the macroscopic scale of momentum $k_0$ and a
certain scale $k_1$ depending on the instruments resolution power. 
Thus, even if one accepts GNS as a model for the motion, the
fluctuation theorem relations would not apply directly to the observed
data.  One needs to know that the local (in momentum space) vorticity
fluctuation, {\it i.e. } the vorticity fluctuations of the amount of vorticity
below scale $k_1$, also obey the fluctuation theorem \equ(2.2).

But this is not a consequence of the theory although one can imagine
further assumptions that would imply such a "local" version of the
theorem (which would apply equally well to the NS equations by the
equivalence conjecture, because the statement would now involve only
quantities local in the above sense).  Therefore rather than entering into
more conjectural territory I prefer to look at the above discussion as
a suggestion of the interest of tests of the mentioned local version of
the fluctuation theorem relations, both in numerical experiments and in
real fluid experiments.
 
{\it A check of \equ(2.2) or \equ(2.4) and of the equivalence
conjectures might be more accessible in the case of fluid
systems (numerical or real, compared to the corresponding conjectures
for interacting particles systems) because they should show chaotic
motions with relatively few degrees of freedom so that the large
fluctuations, that must occur in order to make possible direct testing
of the fluctuation theorems, are more likely to occur and be
observable}.

{\it However } the latter remark also means that the attractors have a
very small size, compared to that of phase space: hence a problem in
the interpretation of \equ(2.2) is the fact that the quantity $\s_0$
that appears in the theorem does coincide with the easily determined
(see \equ(1.10)) contraction of volume in the phase space {\it only if}
the attractor is dense in phase space.
 
This is a property that can be expected in the case of nonequilibrium
statistical mechanics under small or moderately strong forcing, see
[4] for a discussion of this point, but it cannot be
expected at large forcing or in fluid systems.  In the latter case one
should use the contraction rate of the volume elements on the
attractor, see [4], [28].  In the GNS systems it
is likely that the property never really holds.
 
This might render \equ(2.2) quite useless as it is usually unrealistic
to hope to determine the attractor equation with accuracy sufficient to
compute its area contraction per unit time (assuming, as the chaotic
hypothesis implies, that the attractor can be regarded as a smooth
surface).
 
Nevertheless in nonequilibrium statistical mechanics further analysis
is possible, based on an important symmetry of the Lyapunov exponents
spectrum, and one can relate easily the area contraction on the surface
defined by the attractor and the total phase space volume contraction.
Hence one can try to push further the analysis of [4],
[28] to see if one can say something also in the case of the
GNS equations.  \*

\vskip3mm 
{\it\S3 A scenario for experimental checks of the fluctuation theorem.
The Lyapunov spectrum of ED and GED equations. Pairing rule and Axiom C.}
\*
 
What follows is a very {\it heuristic analysis} aimed at giving an
argument for the explicit form that the fluctuation theorem \equ(2.2)
will take in the case of the GED equations (and perhaps by the
above discussion on the equivalence conjecture {\it also} for the ED
equations, see also the last comment in \S7).  No pretention of
mathematical rigor is present and the idea is to illustrate the {\it
simplest possible scenario} that I consider possible and that is
compatible with the small (but {\it not empty} and quite constraining)
set of exact results established elsewhere or below.  The interest is
(apart from the subjective feeling of a certain beauty) that the
discussion suggests experiments and checks that have intrinsic interest
and that do not seem to have yet been considered in the literature.
 
We consider first the case of \equ(1.1) {\it in a domain $\O$ with
obstacles}: in spite of the appearances this is an easier case because
in this case we can imagine forcing the system with a {\it locally
conservative force} which is {\it not} globally conservative, like a
field roughly parallel to one axis and tangent to the obstacles (one
can imagine a uniformly charged fluid under an electromotive constant
electric field).
 
Note that in order to have a non trivial forcing the forcing field must
be non globally conservative: otherwise its effect would be just that
of altering the pressure.
 
The Euler equations can, in general, be regarded as {\it hamiltonian
equations} for a system whose configurations are the diffeomorphisms of
the box $\O$ (in our case a torus with, possibly, a few holes)
containing the fluid: they are not directly in hamiltonian form in the
same sense as the (closely analogous) Euler equations for a rigid body
with a fixed point are not immediately hamiltonian ({\it e.g.} they involve
half the number of actual equations of motion).
 
In this way the GNS or GED equations can be regarded as hamiltonian
equations (approximately so, because the cut--off $K$ destroys this
property) {\it modified} by the action of a non conservative force $\V
g$ and by the gaussian constraint that the total vorticity or the
total energy are constants.
 
Of course we exploit the "slip" ({\it i.e. } no friction) boundary conditions
in order to be able to conclude the hamiltonian nature of the Euler
equations.
 
The phase space will then consist of a space larger than the above
$\FF$, see \equ(1.9): its points $(\V u,\V\d)$ will be (cut--off)
{\it velocity fields} and (cut--off) {\it displacement fields}
describing the positions of the fluid particles with respect to a
reference configuration.  We call this the {\it``full phase space''} of
the equations \equ(1.1).
 
The equations for the displacements will be in all models \equ(1.1):
 
$$\dot{\V\d}=\V u(\V\d,t),\qquad \V\d(\V x,0)=\V\d_0(\V x)\eqno(3.1)$$
which, once $\V u$ is known from \equ(1.3), \equ(1.2), permit us to
compute the physical fluid flow and the positions $\V\d(\V x,t)$ of the
fluid particles that at time $0$ were at the points $\V\d_0(\V x)$, away
from the reference configuration position $\V x\in\O$.
 
In the case in which the $\V u$ verify truncated equations also
\equ(3.1) have to be truncated, for instance by replacing each
$e^{i{\V k}\cdot\V\d(\V x)}$ in $\V u(\V\d,t)$ by its truncated Fourier
expansion.
 
The system motions (describing velocity and displacement fields) can be
regarded as motions with $2D_K$ degrees of freedom where, for instance
in the no obstacles case, $D_K$ is the number of non zero modes $\V k$
with $0<|\V k|<K$ (because each $\V \g_{\V k}$ has two complex
components but $\V \g_{-\V k}=\lis{\V \g}_{\V k}$).  This means that
$4D_K$ coordinates are necessary to describe the motion.
 
Hence there are $4 D_K$ Lyapunov exponents, $2D_K$ from the velocity
equations \equ(1.2) and $2D_K$ from the displacements equations
\equ(3.1).
 
{\it In view of the equivalence conjectures we study the equations GNS
and GED when convenient and the NS or ED when convenient.}
 
Out of the $4D_K$ exponents one has to extract, in the GNS or GED cases,
one exponent that is trivially $0$ because of the conservation of the
dissipation rate and one exponent that is trivially zero and
corresponds to the vector field given by the r.h.s.  of the GNS
equation.  Furthermore in the GNS or GED cases two more vanishing
Lyapunov exponents are associated with other constants of
motion.\footnote{3}{ In the case of the GNS equations the {\it helicity}
$\int\V\o\cdot\V u\,dx$ is a constant of the motion and such is $\int
\V\d^2dx$ for the displacement equations.}
 
The other $2N=4D_K-4$ exponents, or in the ED, NS cases all the
$2N=4D_K$ exponents, can be ordered in two groups the first
containing the first $N$ exponents in decreasing order and the second
the remaining $N$ ones in increasing order.
 
The exponents of the first group are denoted $\l^+_j$, $j=1,\ldots,N$
and the ones in the second group are denoted $\l^-_j$, $j=1,\ldots,N$.
We call the two exponents $(\l^+_j,\l^-_j)$ a {\it pair}.
 
{\it We consider first in detail the ED and GED equations.} In the
above context it seems reasonable that in the full phase space of the
GED and ED equations a {\it pairing rule} holds:
 
$$\fra{\l^+_j+\l^-_j}2= const\eqno(3.2)$$
{\it at least when the forcing is locally conservative as we suppose
from now on unless otherwise stated}.  The value of the constant will
be called the "{\it pairing level}" or "{\it pairing constant}", which
must be $\fra12\langle\,\s\,\rangle_+$, see \equ(1.10).
 
The pairing rule, in fact, formally holds in the present ED case.  One
can easily adapt the proof in [5]: this is discussed in the
Appendix A1.
 
The rule then holds also for GED as a consequence of the conjecture ED.
A direct proof can be made along the lines of the work
[6].  In fact the constraint imposed by the definition of
the multiplier $\a$, \equ(1.7), is a constraint of the type called {\it
isokinetic} in [6] and their proof seems to apply "without
change", although I did not check the details (the appendix A1 should
give the background for such an analysis).
 
In the cases in which \equ(3.2) has been proved, [6],
[5], it holds also in a far {\it stronger} sense: the {\it
local Lyapunov exponents},\footnote{4}{{\it i.e. } the logarithms of the
eigenvalues of $\sqrt{J_t^TJ_t}$, if $J_t$ is the local jacobian matrix
of the evolution operator $V_t$, other than those relative to the
directions of the flow or of the imposed constraints or of other
constants of motion} of which the Lyapunov exponents are the averages,
are paired as in \equ(3.2) to a constant that is $j$ independent but,
of course, is dependent on the point in phase space.  We call this the
{\it strong pairing rule}.  See the final comments.
 
Note that the Lyapunov exponents of the full system can also be easily
divided into {\it velocity exponents}, {\it i.e. } the ones of the GED or ED
equations, and the {\it displacement exponents}, {\it i.e. } the others (which
cannot be measured from the GED or ED evolution alone but require also
\equ(3.1)). In fact if we denote symbolically by $(x,y)$ the pair
$(\V u,\V\d)$ then the jacobian matrix of the equations is described by
a matrix having the form $\pmatrix{A&0\cr B&C\cr}$ where $A,B,C$ are
operators.
 
For a further classification of the exponents we shall think that the
Lyapunov exponents are divided into three classes that we call {\it
viscous}, {\it inertial} and {\it slow}. The following scenario
will be again summarized and enriched in the figure in \S4.
\*
 
\0(1) The slow exponents ("{\it slow pairs}") consist of $M$ pairs of
exponents the largest of which is $\le0$ and it is a velocity exponent
corresponding to slow motions of the velocity field, while the other
(necessarily $<0$) exponent of the pair is a displacement exponent and
corresponds to a fast approach to the stationary state of some of the
displacement variables.
 
\0(2) The viscous exponents ("{\it viscous pairs}") consist of $V$
negative velocity exponents describing the fast approach to the
stationary state of the viscous degrees of freedom of the velocity
field: their paired positive exponents are displacement exponents
associated with chaotic motions of the displacement variables.
 
\0(3) The remaining $2P=N-M-V$ pairs ("{\it inertial pairs}") have one
$>0$ and one $<0$ Lyapunov exponents: $P$ of the {\it pairs} are pairs
of velocity exponents and the $P$ {\it other} pairs are displacement
exponents.  The $P$ pairs of velocity exponents are the only pairs of
exponents of the equations for the velocity field that contain one
positive and one negative element: they describe the gross
characteristics of the chaotic motion on the attractor.  It follows
that the three types of exponents can in principle be {\it uniquely
identified} among the $N$ exponents of the velocity field equations,
see also below.
 
\0The existence of a certain number denoted $P$ above of pairs of
exponents, for the velocity field evolution, that are pairs of
exponents of opposite sign does not follow simply from the fact that we
are collecting togheter pairs containing a $>0$ exponent.  In principle
the $>0$ exponents of the velocity field could be paired with negative
displacement exponents.  We think that it is natural that the $>0$
Lyapunov exponents for the velocity field are paired with $<0$
exponents of the velocity field because we associate such pairs with
the motions on the attractor.  Since the GED equations are reversible
it follows from [28] that {\it if the motions are also
supposed to verify a geometric property called in} {\rm[28]}
{\it Axiom C property} (a simple extension of the paradigm of turbulent
behavior, see [1], that is the Axiom A property) {\it then
there must be an equal number $P$ of positive and negative exponents}
for the restriction of the GED equations to their attractor.  It seems
therefore natural to think that they form $P$ pairs.
 
\0The equality of the number of $>0$ and $<0$ exponents for the motion
on the attractor for the velocity fields is due to the existence, in
reversible Axiom C systems, of a {\it local time reversal} map $i^*$
that transforms the attractor into itself anti-commuting with the time
evolution, even when (and this is the rule in fluid dynamics) the
attractor itself is {\it not} time reversal invariant: see
[28].  We proceed under the assumption that the Axiom C
property is verified: for a complete discussion of the property we must
unfortunately refer to [28].
 
\0{\it In Axiom C systems the time reversal symmetry "cannot be lost"}:
when it is spontaneously broken (because the attractor is not time
reversal invariant) it is replaced by a "weaker" symmetry, good enough
to make "effectively reversible" all the motions on the attractor, a
relation similar to the one in fundamental Physics between $T$ and
$TCP$ (the latter being the "real" time reversal as the first is not a
symmetry of the world we see).
 
\0(4) The other $P$ pairs should consist of displacement exponents
exhibiting a rather symmetric behavior with respect to that of the GED
exponents.  Below we are going to suggest that very similar properties
hold for the NS and GNS equations: in that case this further appealing
symmetry seems compatible with (and in fact it was suggested by) the
data on the velocity Lyapunov spectrum for models ("GOY shell models")
whose behavior is "believed" to be related to NS equations: see
[8], figure in p.  71, taking into account that the
pairing level in such data is very small because the viscosity is very
small.
 
\0In the above scenario the existence of the other $P$ pairs of
displacement exponents is assumed in order to make the total count of
the number of exponents correct and is not based on evidence of any
other kind.  The displacement exponents have been considered in the
literature, [29], but no pairing rule seems to have been
proposed or to have emerged yet (not surprisingly in view of the
difficulty of the measurements).
 
Thus if we measure the Lyapunov exponents for the GED equations alone
we expect to find $P$ pairs of opposite sign exponents paired at the
value $\fra1{2N}\langle\,\s\,\rangle_+$ for some $P\le N$.
 
It seems reasonable that the $P$ pairs of displacement exponents {\it
coincide} with the $P$ pairs of inertial exponents for the velocity
field equations: but this is not really necessary in order that
an unambiguous identification of the three type of exponents be
possible. They are already identified by the above properties.
 
{\it However} if the $P$ pairs of velocity inertial exponents and the
$P$ pairs of displacement inertial exponents do coincide we see that,
by the pairing rule, the knowledge of the Lyapunov spectrum for the
velocity equations {\it implies that all the displacement exponents} are
knwon as well: no need to compute them.
\*
 
\vskip3mm 
\0{\it\S4 Fluctuation theorem predictions for GED and pairing
rule for GED and ED.}
\*
 
With the scenario developed in \S3  we reconsider the fluctuation
theorem and note that it is easy to check, by evaluating the divergence
of the {\it r.h.s.} of equations \equ(1.2), \equ(3.1),
that the volume
in the full ($2N$ dimensional) phase space contracts at the {\it same} rate
$\s$ at which the volume in velocity space does.  Fluid
incompressibility, and absence of displacement variables in the
equations for the velocity field, imply this property.
 
Furthermore if {\it the strong pairing rule} is assumed the
total volume contraction in the full phase space, including the
displacement variables, will be $\s(\V \g)$ and {\it it will be
related} to the contraction $\s_0(\V \g)$ of the area on the attractor
surface by $\s_0(\V \g)=\fra{2P}{2N}\s(\V\g)$, see [4] where
the same mechanism was first exploited.  This gives proportionality
between the "apparent" contraction rate $\s$ and the "true" contraction
rate $\s_0$ on the attractor for the GNS equations.
 
As discussed at the end of \S2 {\it the fluctuation theorem holds for
the fluctuations of} $\s_0$ so that the fluctuations of $\s$ will
verify \equ(2.2) {\it but with a r.h.s} in which $1$ is replaced by
$\fra{P}N$ where $P$ is the number of pairs of Lyapunov exponents for
the GED equations with one positive element.
 
If the number of degrees of freedom is increased by increasing $K$ one
should expect, therefore, that the constant $\fra{P}N\langle\,\s\,\rangle_+$
approaches $P\langle\,\a\,\rangle_+$ because the number of "true exponents" ({\it i.e. }
inertial exponents) {\it should not change} as soon as $K$ is so large that
the motion is well described by the truncated equations: in fact, if
there is ultraviolet stability, the attractor dimension should not
depend on the truncation scale $K$ (as long as it is large enough).
\vfill\eject

Since the conjecture of \S2 implies $\langle\,\a\,\rangle_+=\ch$, the constant
{\it should approach} $P\ch$, at least if $R$ ({\it i.e. } the Reynolds number)
is large. The fluctuation theorem will thus take the form:

$$\fra{\z(p)-\z(-p)}p=P\,\ch\eqno(4.1)$$
if the variable $p$ is defined as in \equ(2.1) {\it but with the
measurable $\s$} replacing the {\it a priori} dif\-ficult to measure
$\s_0$, and $\z(p)$ is the limit of $\t^{-1}\log\p_\t(p)$ with
$\p_\t(p)$ being the $\m_{\e,ged}$ distribution of $p$. The number $P$
is accessible by measurements performed {\it only} on the GED equations
and not involving the displacement variables (being the number of
positive Lyapunov exponents of the GED equations).
 
The above analysis is {\it somewhat conjectural} but experiments, at
least numerical ones, are possible to check the picture; {\it e.g.} one could
attempt at:

\0(1) checking the just derived slope $P\ch$ or
 
\0(2) checking the following picture, representing the above
classification of the exponents:
\vskip1truemm
 
\figini{eqfig1}
\8</punto { gsave >
\8<3 0 360 newpath arc fill stroke grestore} def>
\8</puntino{ gsave 2 0 360 newpath arc fill stroke grestore} def>
\8</origine1assexper2pilacon|P_2-P_1| { >
\8<4 2 roll 2 copy translate exch 4 1 roll sub >
\8<3 1 roll exch sub 2 copy atan rotate 2 copy >
\8<exch 4 1 roll mul 3 1 roll mul add sqrt } def>
\8</punta0{0 0 moveto dup dup 0 exch 2 div lineto 0 >
\8<lineto 0 exch 2 div neg lineto 0 0 lineto fill >
\8<stroke } def>
\8</dirpunta{>
\8<gsave origine1assexper2pilacon|P_2-P_1| >
\8< 0 translate 7 punta0 grestore} def>
\8<>
\8</curva{>
\8</h {0.1} def>
\8<gsave>
\8<3 index 3 index moveto >
\8<4 copy exch 4 1 roll sub neg 3 1 roll sub exch 7 2 roll pop pop >
\8<0 h 1 {dup dup dup 1 sub neg mul 5 index mul exch 6 index mul add 2>
\8<index add exch 6 index mul 3 index add exch lineto } for stroke>
\8<grestore} def>
\8<>
\8</linea {4 2 roll moveto lineto stroke} def>
\8</1linea {1 setlinewidth linea} def>
\8</2linea {2 setlinewidth linea} def>
\8<>
\8<gsave>
\8<>
\8</L {60} def>
\8</LL {30} def>
\8</S {50} def>
\8</H {90} def>
\8<>
\8</OO {0 H} def>
\8</IO {H H} def>
\8</OI {0 2 H mul 40 sub} def>
\8</OM {0 0} def>
\8<>
\8<OO IO dirpunta OM OI dirpunta>
\8<OM moveto OI lineto stroke OO moveto IO lineto stroke>
\8<>
\8</C {5} def 
\8<2 setlinewidth C 0 H 10 sub S 2 div H 0.80 add curva 
\8<S 2 div 2.5 sub H S H 2linea 
\8<0 L S L 1linea>
\8<>
\8<1 setlinewidth>
\8<>
\8</1OO {110 H} def>
\8</1IO {180 H} def>
\8</1OI {110 2 H mul 40 sub} def>
\8</1OM {110 0} def>
\8<1OO 1IO dirpunta 1OM 1OI dirpunta>
\8<1OM moveto 1OI lineto stroke 1OO moveto 1IO lineto stroke>
\8<110 L 150 L 1linea>
\8<1 setlinewidth>
\8</2OO {220 H} def>
\8</2IO {280 H} def>
\8</2OI {220 2 H mul 40 sub} def>
\8</2OM {220 0} def>
\8<2OO 2IO dirpunta 2OM 2OI dirpunta>
\8<2OM moveto 2OI lineto stroke 2OO moveto 2IO lineto stroke>
\8<220 L 280 L 1linea>
\8<>
\8</P0 {110 100} def>
\8</P1 {150 110} def>
\8</M0 {110 20} def>
\8</M1 {150 10} def>
\8<>
\8</Q0 {220 110} def>
\8</Q1 {250 90} def>
\8</Q2 {280 90} def>
\8<>
\8</N0 {220 10} def>
\8</N1 {250 LL} def >
\8</NN1 {250 5 add LL 3 add} def>
\8</N2 {280 LL} def >
\8<>
\8</C {5} def>
\8</CC {10} def>
\8<>
\8<2 setlinewidth>
\8<CC P0 P1 curva>
\8<CC neg M0 M1 curva>
\8<>
\8<C neg N0 NN1 curva N1 N2 2linea>
\8<[3] 0 setdash>
\8<C Q0 Q1 curva Q1 Q2 2linea>
\8<C neg 0 LL 10 add S 2 div LL curva 
\8<S 2 div LL S LL 2linea 
\8<grestore>
\figfin
 
\eqfig{300pt}{150pt}{
\ins{20pt}{140pt}{  $\l$}
\ins{120pt}{ 140pt}{ $\l$}
\ins{220pt}{ 140pt}{ $\l$}
\ins{60pt}{ 80pt}{ $x=\fra{j}{M}$}
\ins{160pt}{ 80pt}{$x=\fra{j}P$}
\ins{250pt}{80pt}{ $x=\fra{j}V$}
\ins{-30pt}{ 60pt}{${\st pairing\atop level}$}
}{eqfig1}{}
 
\0{Fig.1: A sketch of the pairing rule implications on the lagrangian
Lyapunov exponents for the GED equations.}
\*
 
The continuous line in the first graph gives the value ($\le0$) of the
$j$--th (among $M$) slow Lyapunov exponent (as a function of $\fra{j}M$)
of the GED equations; the dashed line is the graph of the paired
exponents (of the displacement equations) and the intermediate line is
the pairing constant.  The exponents are defined only for $x=\fra{j}M$
but the graphs give, instead, continuous (or dashed) lines for visual
aid.
 
The second graph gives the values of the $j$-th (among $P$) pair of
inertial exponents of the GED equation (one positive and one negative
per pair): here too we use the continuous curves even though the number
of such exponents will ususally be much smaller than the total number
and therefore a discrete representation would be more appropriate.
 
The negative curve in the third graph is the graph of the $j$-th
viscous exponent (out of $V$) of the GED equation; the corresponding
positive curve (dashed) is the curve of the companion exponents which
correspond to displacement exponents.  A fourth graph giving the other
$P$ displacement exponents (in pairs of one negative and one positive)
would be qualitatively equal to the second graph (with the curves
dashed for consistence of notation).
 
The graphs {\it are not experimental data}: they are just sketches
illustrating the "simplest" picture that I considered reasonably
possible. They should be taken as a conjecture, and they suggest
performing experimental evaluation of the exponents for a check of the
ideas of the present paper: note that the pairing statement is on a much
firmer footing than the others as it admits a formal proof, described in
the appendix for the ED case.
 
\*
What do we imagine to happen when the equations are changed by enlarging
the cut off (in the velocity as well as in the displacement variables)
or by changing the forcing?  Suppose that the cut off is already so
large that adding one extra mode does not really affect the qualitative
and quantitative features of the motion.  Then adding one mode, {\it i.e. }
increasing the total dimension of the system by $2$, should add one
pair of viscous exponents at the end of the spectrum respectively equal
to $0$ and $-\ch$, as drawn in the third graph in the figure.  While
changing the forcing should, {\it from time to time} as the forcing
changes, change the category of some exponents.  Namely the simplest
picture would be that one of the vanishing slow GED exponents "becomes"
positive and one of the viscous "becomes" inertial and paired with it;
a symmetric evolution should take place with the displacement
exponents.  Or vice-versa.  The attractor changes dimension by $2$
units, see [28],[4], at each of such events.
 
In order that the latter picture be possible one needs that {\it at a
transition} the viscous spectrum bottom consists of a pair of
a $>0$ displacement exponent and a negative viscous exponent
coinciding with the inertial exponents top pair; and that the same
should happen for the bottom pair of the inertial and the top pair of
the slow spectra.
 
The case of periodic boundary conditions does not fit in the above
analysis of the pairing rule because on the torus there is no way of
forcing the system with a locally conservative but globally non
conservative force field with $0$ average.  Nevertheless some kind of
pairing might still occurr under simple non conservative forcing acting
only on some large scale modes, see \S5.
 
It has been pointed out to me by F.  Bonetto that consistency of the
picture {\it requires that the sum of the displacement exponents
be exactly $0$}: the two of us have indeed been able to verify that this
property is formally exactly verified in the ED equations.  And this led
to a correction of the graphs drawn in the figure above that I had
originally drawn without taking such property into account.  We shall
come back on this point in a future study. Note also that the fact that
the sum of the displacement exponents vanishes provides a natural test
that the truncation that one is using is actually large enough for
having reached cut-off independence of the asymptotic properties ot the
motions: this happens at the cut off value where the sum of the
displacements exponents vanishes: further addition of modes only makes
longer the flat part in the third graph of the figure above. \*
\*
 
\vskip3mm 
\0{\it\S5 The NS and GNS equations. Extension of the pairing rule.}
\*
 
We turn to the NS and GNS equations, whose interest is far
greater than the just studied ED or GED equations.
 
One is tempted to say that the scenario should be the same.  However the
pairing rule analysis, which is essential for the physical
interpretation of the results, is no longer naively possible, not even
at a heuristic level.
 
A pairing rule, first pointed out in special {\it non constant}
friction cases in [2], p.  281, has been proved only in the
case of systems subject to special gaussian constraints, see
[6], but it has apparently a much wider validity, see
[2], [30], [7], [31] and
it is likely to hold also in the cases GNS and NS, {\it at least in some
sense}.
 
But the argument in [5] implies the existence of pairing in
systems that are obtained from hamiltonian systems by adding to them an
irreversible constant friction term {\it proportional to the momenta in
a system of canonical coordinates}.  And the argument in [6]
is restricted to "isokinetic" constraints {\it precisely} because they
are reversible constraints that are obtained by adding to a hamiltonian
system a suitable force proportional to the canonical momenta.  Since
this is an essential feature for the validity of \equ(3.2) the latter
becomes doubtful in the cases (that include NS and GNS equations) in
which the friction or thermostat forces are proportional to the
canonical momenta via a matrix $C$ which is not the identity (it is the
laplacian in the case of the NS or GNS equations).\footnote{5}{I owe to
F.  Bonetto the clarification of this point.}
 
In such cases one could envisage that \equ(3.2) is replaced, in the GNS
equations case, by a relation like:
 
$$(\l_j^++\l_j^-)/2=\langle\,\b\,\rangle\,c_j\eqno(5.1)$$
where $\langle\,\b\,\rangle$ is the $\m_{\h,gns}$ average (in the case
of the NS equations one would write $\n$ instead of $\langle\,\b$); and
$c_j$ is some suitable function of $j$, that might be related to the
spectrum of the matrix $C$.  However attempting at establishing such a
connection would lead to too many too detailed assumptions at this stage
and one would like not to rely on them.  And from the proofs in [5], [6]
it seems unlikely that a pairing rule can hold in a strong form, {\it
i.e.} that \equ(5.1) holds for the local exponents if
$\langle\,\b\,\rangle$ is replaced by $\b$.
 
We therefore {\it define} $c_j$ by the \equ(5.1) {\it without linking
$c_j$ to the matrix $C$}.  However we shall suppose that \equ(5.1)
holds in a "almost local" form {\it in the sense that on a rapid time
scale \equ(5.1) becomes true also for the local exponents}.  This means
that, {\it up to an error that tends to zero very quickly with the time
$\t$}, the logarithms of the eigenvalues of the matrix
$(J_\t^T(x)J_\t(x))^{1/2\t}$, with $J_\t(x)$ being the jacobian matrix
for the evolution operator $V_\t$ at $x$, divided by $\t$ verify
$\fra12(\l^+_j+\l^-_j)= c_j \b_\t(x)$ with $\b_\t(x)$ denoting the
average $\fra1\t\int_{-\t/2}^{\t/2}\b(V_tx)\,dt$, still $x$ dependent
because $\t$ is fixed.
 
This property, together with the Axiom C assumption, suffices to
extend, in a suitable form, the validity of the predictions ({\it i.e. }
conjectures) discussed in the previous section for the ED and GED
equations to the case of the NS and GNS equations as follows.
 
We remark that the really relevant feature of the pairing rule,
as far as the applications in [4] and above are concerned,
is not the constancy of the pairing {\it but, rather, the fact that
some kind of pairing takes place on a fast enough time scale}. 
Secondly we assume that this is actually the case for the GNS
equations.  On this remark and on this assumption we base the analysis
of the fluctuation theorem predictions for the GNS (and perhaps, as in
the GED case, for the NS equations).
 
If a relation \equ(5.1) holds the constants $c_j$ will have to add up to
the sum of the Lyapunov exponents The latter can be derived as the
average value of $\s$: this means that $\langle\,\s\,\rangle_+=2\lis
D_K\langle\,\beta\,\rangle_{\h,gns}=(\sum_{j=1}^{2D_K}c_j)
\langle\,\beta\,\rangle_{\h,gns}$, up to terms neglegible as the
Reynolds number tends to $\infty$, see \equ(1.10) and comments preceding
it.
 
Furthermore let $I$ be the set of $P$ inertial pairs ({\it i.e. } of
pairs of Lyapunov exponents $\l^+_j,\l^-_j$ with one positive element)
and suppose that the \equ(5.1) becomes valid on a sufficiently fast time
scale, then the values of $\langle\,\s\,\rangle_+$ and
$\langle\,\s_0\,\rangle_+$ would have ratio (see [4]) $(\sum_{j\in I}
c_j)/(\sum_j c_j)$ so that:

$$\langle\,\s_0\,\rangle=\fra{\sum_{j\in I} c_j}{\sum_j
c_j}\langle\,\sigma\,\rangle_+=(\sum_{j\in I}
c_j)\,\langle\,\b\,\rangle_{\h,gns}= (\sum_{j\in I} c_j)\n\,{\buildrel
def\over =}\,\lis{{P}}\,\n\eqno(5.2)$$
having used the conjecture NS of \S1 equating
$\langle\,\b\,\rangle_{\h,gns}$ to $\n$.
 
Then if a local time reversal exists on the attractor ({\it i.e. } if the
geometric Axiom C is assumed as well, [28], for the dynamics
generated by the GNS equations) the fluctuations of the observable
$\sigma$ will have a "free energy" (or a "generalized sum of Lyapunov
exponents" to adhere to the terminology in [9], [8])
$\z(p)$, in the sense of \equ(4.1), with an odd part $p\,\lis{{P}}\,
\n$, with $\lis{{P}}$ defined in \equ(5.2).  This is a property whose
validity can be conceivably tested in moderately
turbulent GNS systems.  At least the linearity in $p$ of $\z(p)-\z(-p)$
should be observable. For the NS equations the same comments in \S2 and
at the end of \S7 for the ED equations has to be made: the fluctuation
theorem might not hold but a local version of it may hold (see \S2 and
\S7).
 
Note also that, in all cases, the pairing rule is trivially valid in
the case of no forcing: in fact the equivalence criterion in the
conjecture in \S2 requires that in absence of forcing one has to take
$\h=0$ or $\e=0$: {\it i.e. } the stationary state is, in that case, the
trivial (non chaotic) flow $\V u=\V 0, \V\d=const$.
 
The assumption that the forcing be locally conservative {\it has not
been used and disappears} togheter with the constancy of the pairing:
the above more general pairing hypothesis (see \equ(5.1) and the comment
following it) is more "flexible" and does
not require the special hypothesis of local conservativity of the
forcing.  \*
 
\vskip3mm 
\0{\it\S6 Relation between the NS and ED equations. The {\sl barometric}
formula.}
\*
 
Finally we discuss another main point of our analysis.
 
In reference [11] the argument leading to the conjecture NS
above can be interpreted as saying that NS and ED are {\it also} in some
sense equivalent.
 
The argument is based on the constancy of the dissipation rate in a
stationary flow at high Reynolds number and on the {\it microscopic}
reversibility. In some sense the GNS equations emerge as {\it even more
natural} than the NS equations.
 
A criticism can be raised, however. In fact one can argue that the
energy is {\it also} constant in a stationary state and one could
develop the argument in [11] to imply that the GED equations
are also a good model for a fluid motion.
 
Since clearly one should not expect NS and ED to be equivalent this
looks at first as an unsoluble logical contradiction. Which can
furthermore be conceivably easily checked to occurr.
 
However on further thought the contradiction can be resolved and one
should think that all what has been deduced is that there should be a
relation between the stationary states of ED (or GED equivalently) and
of NS (or GNS). The relation to which I think is the kind of relation
that one also finds in equilibrium statistical mechanics in
gases in a strongly varying external field of intensity $g$, like the
gravity field.
 
{\it Locally} a gas in a field looks just like a homogeneous gas in
equilibrium, but globally over a length scale $H$ over which the
external potential really changes ($\b m g H\sim 1$, if $\b$ is the
inverse temperature and $m$ the particles mass) one will see that
pressure and the density are not constants and one gets the {\it
barometric formula}, see [32].
 
Likewise we can expect that the stationary states of ED (or equivalently
of GED) are {\it also} ``locally'' equivalent to stationary states for
NS (or GNS): in the sense that if we only look at observables depending
on field components $\V u_{\V k}$ with modes ${\V k}$ on a certain scale
$|{\V k}|\sim\k$ whose size depends on the dissipation then we should see
essentially no difference. The precise relation that determines $\k$
will be called {\it barometric formula}:
it should be easy to determine the formula on the basis of
dimensional considerations.  {\it Locality} is here to be interpreted in
momentum space rather than in coordinate space.
 
The determination of the barometric formula amounts essentially at a
development of the analogous of the Kolmogorov theory for the ED
equations.
 
We now attempt at a {\it partial} development of such theory, in the no
obstacles case for simplicity, on the basis of a few assumptions that
deserve further attention and perhaps criticism.
We follow
closely the ideas (and imitate the assumptions) of the exposition of
Kolmogorov's theory in [14].  We set $\r=1$.
 
It seems reasonable to suppose that in the ED case the stationary
distribution equipartitions the energy among the modes,
{\it i.e. } $\langle\,{|\V\g_\V k\,\rangle|^2}=\g^2$ for all $\V k$ 
in the "inertial range"
$L^{-1}\ll |{\V k}|\ll k_\ch$ where $k_\ch$ is the "Kolmogorov scale", to
be determined below. Hence $\g^2(k_\ch L)^3=\e$.
 
Then a velocity variation characteristic of the momentum scale $\k$ is
given by the expression $v_\k^2=\langle\,{(\sum_{|\V
k\,\rangle|\in[\k/2,\k]} \V\g_{\V k})^2}$ and, assuming statistical
independence of the distribution of the various $\V\g_{\V k}$, we get
$v_\k^2=(\k L)^3\g^2$ up to a constant factor.
 
The cut-off scale $k_\ch$ has to be, on dimensional basis, a momentum
scale formed with the quantities $k_0, \sqrt{\ch^2/\e}$ and
$\ch/\sqrt{gL}$ or, in case of ultraviolet instability of the equations
solutions, it might even depend on the cut off $K$ necessary to make
the equations well defined.  Hence it cannot be determined without a
more detailed theory of the equations.
 
For purposes of comparison we note that the quantity called $\e$ in the
Kolmogorov's theory ("K41-theory"), see [14], corresponds to
$\h\n$ of the present paper.
 
In this case the energy distribution ({\it i.e. } the amount $K(k)dk$ of energy
per unit volume and between $k$ and $k+dk$) is
$K(k)=\fra{3\e}{4\p}\fra{k^2}{k_\ch^3}$, for $k<k_\ch$: very different
from the Kolmogorov's $k^{-5/3}$ law.
 
In the K-41 theory a key role is plaid by the quantity $v_\k^3\k$ which
is {\it identical to $\h\n$ for all $k_0\ll \k\ll k_\n$}. Therefore we
compute the value of $v_\k^3\k$ in our case and we find:
 
$$\fra{v_\k^3\k}{\e\ch}=\fra{((\k L)^3\g^2)^{3/2}\k}{\e\ch}=
\fra{\big((k_\ch
L)^3\g^2\big)^{3/2}k_\ch}{\e\ch}(\fra{\k}{k_\ch})^{11/2}=
\fra{\e^{3/2}k_\ch}{\e\ch}\big(\fra{\k}{k_\ch}\big)^{11/2}\eqno(6.1)$$
and we see that the quantity $v_\k^3\k$ does depend on $\k$ in the ED
case.  

Given $\k$ the SRB statistics for the ED equations driven with a
total energy $\e$ gives to this quantity the same value that it has in
the SRB statistics for the NS equation driven with a total vorticity
$\h$ if:
 
$$\fra{\e\ch}{\h\n}=const \,\k^{-\fra{11}2}\eqno(6.2)$$
provided (of course) $\k$ is greater than the Kolmogorov scales
$k_\n,k_\ch$: since the constant depends on $k_\ch$ the above
discussione leads only to the determination of the exponent $11/2$.
 
The ``barometric formula'' is then the statement of equivalence
between NS and ED {\it on scale} $\k$, {\it i.e. } if one only looks at field
properties depending on $\V \g_{\V k}$ for $\fra12
\k<|{\V k}|<\k$, if \equ(6.2) holds and $\k\gg k_\n,k_\ch$.
 
If we look at a different scale $\k'=2^n\k$ for some (large) $n$ then we
can expect equivalence between ED (or GED) and NS (or GNS) {\it but} the
pairs $\e,\h$ should now be such that \equ(6.2) holds on the new scale:
the analogy with the usual barometric formula for the Boltzmann Gibbs
distribution in the gravity field justifies the name given to \equ(6.2).
We see that $\h\n$ plays the role of the gravity, $\e\n$ plays the role
of the chemical potential and $\k/k_\ch$ plays the role of the height.
 
The above analysis seems to be fully consistent with the numerical
results in [12] who first proposed, in a different context, a
picture very close to the one developed here.
 
It is clear that this point of view has several consequences: for
instance in particular it tells us that that the shape $j\to c_j$
of the pairing curve in \equ(5.1) cannot be arbitrary ({\it i.e. } $\langle\,\b\,\rangle
c_j\sim \n {\V k}_j^2$ if the modes are ordered in increasing order).
This is a point on which I hope to return in a later analysis.
 
Also: the equivalence between NS and ED on a given momentum scale makes
more interesting the ideas in [23] and a test of the Onsager reciprocity
derived in the latter paper seems now quite feasible and seems also to
have conseqiences for the real NS equations.
 
A further remark is that although \equ(6.2) depends on the validity of
the K-41 theory and of the corresponding theory for ED equations the
barometric formula can be developed {\it independently} of such
theories: hence any modification of the K-41 theory (and of the
corresponding theory for ED) will lead to a barometric formula, with
a relation between $\e,\h,\k$, possibly more complicated than \equ(6.2).
 
It should be remarked that the above analysis is based on the
conjectures NS and ED in \S2 ({\it i.e. } it is independent on the chaotic
hypothesis) and the observables involved are observables relative to a
fixed shell, so that they are "local" in the sense discussed in \S2,
after conjecture ED.  This is necessary as the conjectures (may) fail
for (some) non local obserbables, see \S2.

\*
\vskip3mm 
\0{\it\S7 Overview and concluding remarks.}
\*
 
1) The analysis is based on Ruelle's proposal of considering the SRB
distributions as the "physical distributions" describing stationary
states of generic mechanical systems, {\it i.e} systems with at least
one positive Lyapunov exponent: technically this is formulated by
assuming the validity of the "chaotic hypothesis" of \S2,
[10].  Even in equilibrium this hypothesis is stronger than
the ergodic hypothesis: therefore there is no hope to prove it in "any"
system.  But one can analyze its consequences, just as in the case of
the ergodic hypothesis.  

The consequences should be something like non equilibrium
thermodynamics and turbulence theory: they can be expected to be of
general nature (as Boltzmann's heat theorem, {\it i.e. } the exactness of
$(dU+pdV)/T$).  Noting that the same system can be described
equivalently by several different equations ({\it e.g.} a fluid can be
described by microscopic equations in terms of its molecules or by
equations for a macroscopic continuoum) one may expect that the
stationary states can be described by several probability
distributions.  In [11] the analogy between this obvious
remark and the possibility of many equivalent ensembles in equilibrium
statistical mechanics was pointed out (with reference to fluids but the
same ideas apply to molecular models).  Here I tried to give an example
of how one can pursue the analogy to build the notion of "equivalent
dynamical ensembles": the idea seems to be slowly emerging,
independently, in various areas and the first attempt can be identified
in [12].
 
2) The mechanism for the generation of equivalent ensembles ({\it i.e. } SRB
distributions if one follows Ruelle) is illustrated by considering the
two pairs of equations in \equ(1.1).  In [11] there is a
heuristic argument of why one can consider the first two equations (NS
and GNS) equivalent: conjecture NS above.  The second conjecture is
based on the same type of reasoning.  Basically I state that the time
scale over which the variable coefficient $\b$ or $\a$ in \equ(1.1)
reaches its average value (respectively equal to the viscosities $\n$
and $\ch$) should be much faster than the time scales for the
hydrodynamic evolution of the NS or ED equations.
 
3) The equivalence conjecture leads to studying the stationary ensembles
in the most convenient case, which may be NS (respectively ED) or GNS
(respectively ED) depending on which property one studies.  Just as in
the case of statistical mechanics it is sometimes more convenient to
study the canonical rather than the microcanonical ensemble ({\it e.g.}
Boltzmann's heat theorem is very conveniently studied in the canonical
ensemble while entropy theory is more natural in the microcanonical
ensemble).
 
The main application of the chaotic hypothesis is the "fluctuation
theorem" (\equ(2.4)) which is a rigorous consequence of the chaotic
hypothesis (a mathematically minded treatment of the original proof
given in [10] can be found in the third of [20]
for the case of maps or in [26] \0for flows). 

One does not know whether the GNS or GED equations verify the
hypothesis: therefore it is interesting to test the consequence
\equ(2.4) of the fluctuation theorem. Its verification would provide
stronger grounds for the hypothesis because it is an {\it exact
parameterless prediction} (to use an ambitious comparison one can say
that it is as an exact consequence of the chaotic hypothesis as
Boltzmann's heat theorem is of the ergodic hypothesis: both hyptotheses
are clearly violated in interesting case, but al least the heat theorem
is always valid).
 
4) Checking the fluctuation theorem result for the GED or GNS equations
meets a major obstacle: namely (by the equivalence conjecture) the GED
or GNS systems are "far from equilibrium".  This means that the
stationary distributions will live on small attracting and basically
unknown sets ({\it i.e. } with closure smaller than the whole phase space and
possibly fractal).  This difficulty was already met in experimental
attempts at checking the chaotic hypothesis in the case of simple gas
conduction models in [4].

In that case the difficulty was not really severe as the attractor
dimension was almost maximal, but it stimulated a proposal for a
general solution based on a very remarkable property of the Lyapunov
exponents in certain dissipative systems.  This property is rigorously
established for systems which are locally hamiltonian ({\it e.g.} forced by an
electromotive force) and subject to a viscous force proportional to the
momenta, or alternatively subject to a constraint of constant kinetic
energy ("isokinetic" systems) imposed via Gauss' minimal constraint
principle.  The result is that the Lyapunov exponents can be arranged
in pairs, {\it even the local ones}, with constant average
[5], [6].  In close analogy with the conservative
cases, where they can be arranged in pairs with {\it zero} average. 
The average is the constant friction coefficient in the first case and
the Gauss' multiplier or its time average in the second case (depending
on whether one considers the local or average Lyapunov exponents).
 
5) In this paper I have shown that the GED equations and the ED
equations fall, {\it respectively}, under the assumptions of the
mentioned papers except that the systems here have infinite dimension:
this is achieved by considering the {\it full} lagrangian equations for
the fluid motion (the theorem does not hold if one only looks at the
velocities of the fluid, {\it i.e. } unless the phase space is enlarged to take
into account the displacements of the fluid elements).  The average of
the pairs is equal to the viscosity $\ch$ in the ED case and, in the
GED case, is equal to the average viscosity $\langle\,\a\,\rangle$ or (for the
local exponents) to the viscosity $\a$, see \equ(1.1)),
 
The infinite dimensionality makes the result formal and one can
think that it could be made rigorous if one could establish ultraviolet
stability for the ED or GED equations ({\it i.e. } that the equations can be
cut off in momentum space at a large enough, Reynolds number dependent,
cut--off): but this is {\it out of question} in $3D$, as one does not even
know the corresponding property for the NS equation, [15]. 
In the $2D$ case (not dicussed here) there is more hope: even the Euler
equations in this case do not have ultraviolet problems ({\it i.e. } smooth
data evolve into snmooth solutions).  But unfortunately the existence
and regularity theorems for the Euler equations {\it are not
constructive} (at least I do not know of any proof that does not use a
compactness argument or a monotonicity argument) and I see no hope to
tackle the question until a constructive theory of $2D$ Euler equation
is available (I find it surprsing that this seems hardly considered a
problem and the $2D$ Euler existence and uniqueness theory is
considered "completely understood").
 
6) Once pairing is (formally) established one can have recourse to the
ideas in [4] which provide a simple relation between the
phase space contraction in the full phase space and the one on the
(unknown) attractor: a property that permits us to put the fluctuation
theorem in a simple and {\it usable} form, \equ(4.1), {\it provided one
can show that the motion on the attractor is reversible}.
 
Of course even in the case of the GED equations that are reversible the
motion on the attractor {\it will not be reversible} (because time
reversal transforms the attractor into a repeller).  The problem has
been studied in [28]: where we investigated under which general
geometric conditions one could establish that {\it time reversal is an
undestructible symmetry}, {\it i.e. } the conditions under which even when time
reversal symmetry $i$ is "spontaneously broken" (as one can intepret
the formation of an attractor smaller than phase space) still one can
define a {\it new} transformation $i^*$ acting on the attractor only
and reversing the time ({\it i.e. } $i$ and $i^*$ are related in the same way
as $T$ and $TCP$ in fundamental Physics).  In [28] it is shown
that a very natural and simple geometric property exists which has
precisely the above feature of rendering time reversal undestructible.
 
The property is a global form of the Axiom A property: the latter is a
property that seems to be rather widely accepted as closely related to
chaotic systems, [33].  A related property is the Axiom B:
which is a global version of the Axiom A (which should be regarded as a
property only of the attractor and not of the whole phase space).  The
Axiom B, particularly if one inteprets literally the original
definition (by Smale, [34]), is not exactly what is needed (and
it seems to be not even structurally stable, unlike the Axiom A
property).  In [28] we show that a "minor modification" of
Axiom B, that we call Axiom C, makes (as a mathematical theorem), time
reversal undestructible in the above sense; furthermore as conjectured
in [28], and as checked recently, Axiom C systems are
structurally stable.
 
Hence the fluctuation theorem applies to GED if: i) the chaotic
hypothesis is strengthened (and simplified) by saying that "chaotic
reversible systems are Axiom C sytems for the purposes ..." (see \S2),
ii) the pairing rule, \equ(3.2) {\it proved} here formally, is assumed
valid and iii) the heuristic argument, given in [4] (and necessary to
intepret the experimental results of that work), connecting the total
phase space contraction to the one on the attractor is assumed.  It
takes the form of \equ(4.1).  Note that the three properties above are
independent and, unless one dismisses formal proofs (the formality being
the application of a finite dimensional argument to an infinite
dimensional case), i) and ii) are based on mathematical theorems.
 
7) Independently on the applications of the chaotic hypothesis to GED,
the pairing rule for the GED and ED equations establishes a {\it very
strong and remarkable} connection between the Lyapunov exponents of the
velocity field evolution and the ones of the lagrangian description of
the same fluid (twice as many).  In \S4 we try to spell out this
relation: the picture that emerges is very appealing, but the reader is
warned that it is only the "simplest" possible consistent with the
pairing rule.  It is not easy to think of others but I am afraid that
there might be others.  This analysis is {\it independent} of the
chaotic hypothesis.
 
8) To apply the above discussion to the GNS (reversible) equations a
new difficulty arises: namely I feel that it is unreasonable to even
think that the pairing rule holds for the NS or GNS equations ({\it i.e. } if
one has a friction proportional to the laplacian of the field or a
gaussian "isovorticity" constraint).  But I propose that a more general
pairing rule holds, \equ(5.1), essentially on the basis that the time
scale for the stabilization of the average of paired Lyapunov exponents
is shorter than the other time scales involved.  This is my suggestion
for the extension of the pairing rule to non isokinetic systems: it is
proposed as the "minimal" assumption necessary to deduce a testable
form of the fluctuation theorem.  It would be interesting to try to
check its validity (or simply the validity of the fluctuation theorem
for GNS).
 
9) Finally partly inspired by the ideas in [12] I try to
establish a relation between NS and ED equations: this requires a
formulation of the Kolmogorov theory for the ED equations.  It is easy
to see that such a theory risks to be controversial.  Hence I refrain
from formulating it in complete form and I only assume that there is an
"inertial range" where the energy is {\it equipartitioned} among the
modes.  This Leads to a scaling law, that I have called {\it barometric
formula} in \S6, wich is relevant in the comparison between the NS or
GNS equations and the ED and GED equations, as a consequence of the
conjectures NS and ED.  This is independent on the Chaotic hypothesis.
\*

As a concluding remark I point out that while the fluctuation theorem
cannot be applied to the NS equations but only to the GNS equations
(because it involves "nonlocal" observables, so that the equivalence
conjecture might be stretched too much) one can still think of applying
it in a local version: this is in analogy with the inapplicability of
the gaussian fluctuation theory to the energy distribution, in
equilibrium statistical mechanics, in the microcanonical ensemble.  The
energy distribution is gaussian in the canonical ensemble ("central
limit theorem") and a delta function in the microcanonical ensemble.
Nevertheless the fluctuations of energy in both ensembles are gaussian
(and equal) if one looks at the energy in a local region ({\it i.e. }
small compared to the total volume).  I think that a similar picture
holds for the fluctuation theorem but this requires a longer discussion
on which I hope to come back in the future, because it would open the
possibility of checking the fluctuation theorem in many more
experiments.\footnote{6}{For the sake of clarity about what I have in
mind by a possible "local version" of the fluctuation relations for NS
or GNS equations I give an example.  Let $k_0$ be the momentum scale of
the container and $k_1$ be a fixed higher momentum scale
($\n$--independent).  Then one can define the entropy production in the
given scale range to be $\s_{k_1}=2 \lis D_{k_1} \b_{k_1}$, where $\lis
D_{k_1}$ is twice the number of modes below $k_1$ (see \equ(1.10)) and
$\b_{k_1}$ is defined by \equ(1.2) with the fields $\V u,\V \o$
truncated at momentum $k_1$.  The distribution $\p_\t(p)$ of the
quantity $p$ (defined as in \equ(2.1) but with $\s_{k_1}$ replacing
$\s$) generates a function $\z(p)$.  A "local fluctuation relation"
could then be \equ(4.1) with $P\ch$ replaced by $\lis P \n$, see
\equ(5.2).\vfil}
 
\*
\0{\it Acknowledgements:} I am grateful to L.  Biferale, G.  Paladin
and A.  Vulpiani for the information about the GOY model spectral
symmetry and to G.  Gentile, C.  Liverani, V.  Mastropietro and
particularly to F.  Bonetto for many clarifying discussions and
suggestions: his criticism has been very important for correcting slips
and improving various points. The paper improved substantially also
by the referee's remarks. This work is part of the research program
of the European Network on: "Stability and Universality in Classical
Mechanics", \# ERBCHRXCT940460 and it has been partially supported also
by MRS grants \#40\%,60\%, and INFN.

\*
\vskip3mm 
\0{\it Appendix A1: The hamiltonian formalism for Euler equations and
Dressler's theorem for ED.}
\*
 
To check the applicability of the results on pairing of [5]
to the ED equations we must check that the equations can be written, in
canonical coordinates for some hamiltonian function $H$, in the form:
 
$$\eqalign{
\dot{\V q}=&{{\V\partial}}_{\V p} H\cr
\dot{\V p}=&-{{\V\partial}}_{\V q}H+\V F-\ch \V p\cr}\eqno(A1.1)$$
where $\V F$ is such that $\partial_{q_j} F_i=\partial_i F_j$ without being
necessarily $\V F=-{{\V\partial}} V$ for some {\it globally defined} $V$ (the
latter would be a trivial case).  The labels for the components of $\V
q$ are $\V x,i$ with $\V x\in \O$ and $i=1,2,3$.  The partial
derivatives are, correspondingly, functional derivatives; we shall
ignore this because a "formally proper" analysis is easy and leads to
the same results.  By "formal" we do not mean rigorous, but {\it only}
rigorous if the functions we consider have suitably strong smoothness
properties: a fully rigorous treatment is of course impossible for want
of reasonable existence, uniqueness and regularity theorems for the
Euler equations or the Navier Stkes equations in $3$ dimensions.
 
Consider first the Euler equations. They can be derived from the
Lagrangian:
 
$$\LL_0(\dot{\V\d}, \V\d)=\fra{\r}2\int\dot{\V \d}^2 d\V x\eqno(A1.2)$$
($\r=$ density) {\it defined on the space $\DD$ of the diffeomorphisms
$\V x\to\V\d(\V x)$ of the box $\O$}, by imposing the {\it ideal}
constraint:
 
$$\det J\noindent{}=\det \fra{\partial \V \d}{\partial\V x}(\V x)=
{{\V\partial}}\d_1\wedge{{\V\partial}}\d_2\cdot{{\V\partial}}\d_3
\noindent{}=1\eqno(A1.3)$$
In fact, if $Q(\V x)$ is a Lagrange multiplier, the stationarity condition
for:
$$\LL(\dot{\V\d},\V\d)=\fra\r2\int\dot{\V\d}^2d\V x+ \int Q(\V x(\V\d))\det
J(\V \d(\V x))\,d\V x\eqno(A1.4)$$
leads to, using $J(\V \d(\V x))d\V x=d\V\d$ :
%
$$\r\ddot{\V\d}=- {{\V\partial}}_{\V\d} Q\eqno(A1.5)$$
so that setting $\V u(\V\d(\V x))=\dot{\V\d}(\V x)$, $p(\V \d)=
Q(\V x(\V \d))$ if $\V\d=\V\d(\V x)$, we see that:
$$\fra{d \V u}{dt}=-\fra1\r{{\V\partial}} p\eqno(A1.6)$$
which are the Euler equations. And the multiplier $Q(\V x)$ can be
computed as:

$$Q(\V x(\V\d))=p(\V\d)=-\big[\D^{-1}({{\V\partial}} \W u\cdot\W\partial\,\V u)\big]_{\V\d}
\eqno(A1.7)$$
where the functions in square brackets are regarded as functions of the
variable $\V\d$ and the differential operators also differentiate over
such variable. After the computation the variable $\V\d$ has to be set
equal to $\V\d(\V x)$.
 
Therefore by using the Lagrangian:

$$\LL_i(\dot{\V \d},\V\d)=\int\big(
\fra{\r \dot{\V\d}(\V x)^2}2-
\big[\D^{-1}({{\V\partial}} \W u\cdot\W\partial\,\V u)\big]_{\V\d(\V x)}
(\det J(\V\d)|_{\V x}-1)\big)\,d\V x\eqno(A1.8)$$
we generate Lagrangian equations for which the "surface" $\Si$ of the
{\it incompressible diffeomorphisms} in the space $\DD$ is {\it
invariant}: these are the diffeomorphisms $\V x\to\V \d(\V x)$ such that
$J(\V \d)={{\V\partial}}\d_1\wedge{{\V\partial}}\d_2\cdot{{\V\partial}}
\d_3\noindent{}=1$ at every point $\V x\in\O$.
 
Then $\Si$ is invariant in the sense that the solution to the
Lagrangian equations with initial data "on $\Si$", {\it i.e. } such that
$\V\d\in\Si$ and ${{\V\partial}}\cdot\dot{\V\d}(\V x)=0$, evolve remaining "on
$\Si$".
 
The Hamiltonian for the Lagrangian \equ(A1.8) is obtained by computing
the canonical momentum $\V \p(\V x)$ and the Hamiltonian as:
 
$$\eqalign{
\V \p(\V x)=&
\fra{\d\LL_i}{\d\,\V{{\dot\d}}(\V x)}
=\r\dot{\V \d}(\V x)+\ldots\cr
H(\V \p,\V q)=&{1\over 2}
\big( G(\V q)\V \p,\V \p\big)\cr}\eqno(A1.9)$$
where $G(\V q)$ is a suitable quadratic form that can be read directly
from \equ(A1.8) (but it has a somewhat involved expression of no
interest for us), and the $\ldots$ (that can also be read from
\equ(A1.8)) {\it are terms that vanish if} $\V\d\in\Si$ and
${{\V\partial}}\cdot \dot{\V\d}=0$, {\it i.e. } they vanish on the
incompressible motions.
 
The above is well known and shows that the Euler flow can be interpreted
as a gedodesic flow on the surface $\Si$ of the incompressible
diffeomorphisms of the box $\O$ enclosing the fluid, see appendix 2
in [35].
 
Modifying the Euler equations by the addition of a force $\V f(\V x)$
such that {\it locally} $\V f(\V x)=-{{\V\partial}} \F(\V x)$ means
modifying the equations into:
 
$$\fra{d\V u}{dt}=-{{\V\partial}} p-{{\V\partial}}_{\V x}\F\eqno(A1.10)$$
which can be derived from a lagrangian:
 
$$\LL^\F_i(\dot{\V\d},\V\d))=\LL_i(\dot{\V\d},\V\d))-\int \F(\V\d(\V x))\,
d\V x\eqno(A1.11)$$
which leads to the equations:

$$\dot{\V u}(\V\d(\V x))=-\fra1\r{{\V\partial}}_{\V\d} p(\V\d(\V
x))+{{\V\partial}}_{\V\d}\F(\V\d(\V x))\eqno(A1.12)$$
 
Hence the ED equations have the form:
 
$$\eqalign{
\dot{\V q}=&\partial_{\V p} H\cr
\dot{\V p}=&-\partial_{\V q} H-\V F-\ch\V p\cr}\eqno(A1.13)$$
at least as far as the motions which have an incompessible initial datum
are concerned. This is true because the ED equations which have an
incompressible initial datum evolve it by keeping it incompressible.
 
The Lyapunov exponents of the equation \equ(A1.13) verify the pairing
rule by the analysis in [5]. However the pairing takes place in
the {\it full} phase space of the diffeomorphisms of $\O$, including the
incompressble ones.
 
It is not difficult to see, by using that the constraint to stay
on the surface $\Si$ is {\it holonomous}, that one can find canonical
coordinates $\V\p,\V\k$, $\V\p^\perp,\V\k^\perp$ describing the motions
on $\Si$ or, respectively, transversally to it. And the equations for
$\V\p^\perp,\V\k^\perp$ are, {\it near} $\Si$ and for $\V\p^\perp$ small,
$\dot{\V \p}^\perp=-\ch \V\p^\perp$ and $\dot{\V \k}^\perp=\V\p^\perp$
so that the corresponding Lyapunov exponents are trivially paired in
pairs $0,-\ch$ with pairing sum $-\fra\ch2$.
 
Since we have seen above that {\it all the exponents are paired} at the
level $\fra\ch2$ this means that {\it all the physically interesting
exponents} (relative to the incompressible motions, {\it i.e. } relative to the
$\V\p,\V\k$ coordinates) are {\it also} paired at the same level, as
claimed in \S4.
\*\*

\centerline{\bf Erratum}
In Sec. 3 the statement 
\*

\0``The pairing rule, in fact, formally holds in the
present ED case at least when the forcing is locally conservative
...''
\*

\0is an error as a ``locally conservative forcing'' cannot admit a periodic
potential, hence any supposed pairing rule cannot be based on
such assumption. Therefore the analysis of Sec. 3 can only be regarded as
speculative assuming an approximate pairing (nevertheless
evidence for the existence of some kind of pairing is present in
the literature).
\*

Sec. 1,2,6 present the main idea of equivalent evolutions and do
not need any pairing property; Sec. 4,5 remain valid if an
approximate pairing rule is supposed; but the claims on its
formal validity for ED and GED (stated in Sec.3 and appendix A1
and mentioned in items 5,7 in Sec.7 ), have to be retracted as a
consequence of the error in Sec.3.
 
\*
\vskip3mm 
\0{\bf References.}

\*
\def\aA{ Arnold, V.: {\sl M\'ethodes Math\'ematiques de la M\'ecanique
classique}, Ed. MIR, Moscow, 1974.}
\rif{A}{}{\aA}{0}
 
\def\aBG{ Bonetto, F., Gallavotti, G.: {\it Reversibility,
coarse graining and the chaoticity principle}, preprint
in {\it mp$\_$arc@ math.utexas.edu}, \#96-50.}
\rif{BG}{}{\aBG}{0}
 
\def\aBGG{ Bonetto, F., Gallavotti, G., Garrido, P.: {\it Chaotic
principle: an experimental test}, in {\it mp$\_$arc@ math. utexas. edu}
\# 154, 1996.}
\rif{BGG}{}{\aBGG}{0}
 
\def\aBJPV{Bohr, T., Jensen M.H., Paladin, G., Vulpiani, A.:
{\it Dynamical systems approach to turbulence}, Cambridge Nonlinear
series, 1996.}
\rif{BJPV}{}{\aBJPV}{0}
 
\def\aDPH{ Dellago, C., Posch, H., Hoover, W.: {\it Lyapunov instability in
system of hard disks in equilibrium and non-equilibrium steady states},
preprint, Institut f\"ur Experimentalphysik, Universit\"at Wien,
Austria, August 1995.}
\rif{DPH}{}{\aDPH}{0}
 
\def\aDr{ Dressler, U.: {\it Symmetry property of the Lyapunov
exponents of a class of dissipative dynamical systems with viscous
damping}, Physical Review, {\bf 38A}, 2103--2109, 1988.}
\rif{Dr}{}{\aDr}{0}
 
\def\aDM{ Dettman, C.P., Morriss, G.P.: {\it Proof of conjugate pairing
for an isokinetic thermostat}, N.S. Wales U., Sydney 2052, preprint,
1995.}
\rif{DM}{}{\aDM}{0}
 
\def\aECM{ Evans, D.J.,Cohen, E.G.D., Morriss, G.P.: {\it Viscosity of a
simple fluid from its maximal Lyapunov exponents}, Physical Review, {\bf
42A}, 5990--\-5997, 1990.}
\rif{ECM}{1}{\aECM}{0}
 
\def\bECM{ Evans, D.J.,Cohen, E.G.D., Morriss, G.P.: {\it Probability
of second law violations in shearing steady flows}, Physical Review
Letters, {\bf 71}, 2401--2404, 1993.}
\rif{ECM}{2}{\bECM}{0}
 
\def\aEM{ Evans, D.J., Morriss, G.P.: {\it Statistical
mechanics of nonequilibrium liquids}, Academic Press, 1990.}
\rif{EM}{}{\aEM}{0}

\def\aER{ Eckmann, J.P., Ruelle, D.: {\it Ergodic theory of strange
attractors}, Reviews of Modern Physics, {\bf 57}, 617--656, 1985.}
\rif{ER}{}{\aER}{0}
 
\def\aFP{ Frisch, U., Parisi, G.: in {\sl Turbulence and predictability},
p. 84--87, ed. by M. Ghil, R. Benzi, G. Parisi, North Holland, 1984.}
\rif{FP}{}{\aFP}{0}
 
\def\aG{ Galilei, G.: "Discorsi intorno a due scienze nuove",
"Opere", Salani, Firenze, 1964.}
\rif{G}{}{\aG}{0}
 
\def\aGa{ Gallavotti, G.: {\it Some rigorous results about 3D Navier
Stokes}, Les Houches 1992 NATO-ASI meeting on ``Turbulence in spatially
extended systems'', ed. R. Benzi, C. Basdevant, S. Ciliberto;
p. 45--81; Nova Science, NY, 1993.}
\rif{Ga}{1}{\aGa}{0}
 
\def\bGa{ Gallavotti, G.: {\it Ergodicity, ensembles,
irreversibility in Boltzmann and beyond}, Journal of Statistical
Physics, {\bf 78}, 1571--1589, 1995. See also {\it Topics in chaotic
dynamics}, Lectures at the Granada school, ed. Garrido--Marro, Lecture
Notes in Physics, {\bf 448}, 1995. And, for mathematical details, {\it
Reversible Anosov diffeomorphisms and large deviations.},
Ma\-the\-ma\-ti\-cal Physics Electronic Journal, {\bf 1}, (1), 1995,
(http:// www. ma. utexas. edu/ MPEJ/ mpej.htlm).}
\rif{Ga}{2}{\bGa}{0}
 
\def\cGa{ Gallavotti, G.:{\it Chaotic hypothesis: Onsager reciprocity
and fluctuation--dis\-si\-pa\-tion theorem}, Journal of
Statistical Physics, {\bf 84}, 899--926, 1996.}
\rif{Ga}{3}{\cGa}{0}
 
\def\dGa{ Gallavotti, G.: {\it Chaotic principle: some applications
to developed turbulence}, archived in {\it mp$\_$arc@ math.utexas.edu},
\#95-232, and {\it chao-dyn@ xyz.lanl.gov}, \#9505013.
In print in Journal of Statistical Physics, 1996.}
\rif{Ga}{4}{\dGa}{0}
 
\def\eGa{ Gallavotti, G.: {\it Extension of Onsager's reciprocity to
large fields and the chao\-tic hypothesis}, mp$\_$ arc 96-109; or
chao-dyn 9603003, to appear in Physical review Letters.}
\rif{Ga}{5}{\eGa}{0}
 
\def\fGa{ Gallavotti, G.: {\it Equivalence of dynamical ensembles
and Navier Stokes equations}, mp$\_$ arc 96-131; or chao-dyn
9604006, to appear in Physics Letters A.}
\rif{Ga}{6}{\fGa}{0}
 
\def\aGe{ Gentile, G.: {\it Large deviation rule for Anosov flows},
preprint http://chimera. roma1. infn. it, 1996.}
\rif{Ge}{}{\aGe}{0}
 
\def\aGC{ Gallavotti, G., Cohen, E.G.D.: {\it Dynamical ensembles in
nonequilibrium statistical mechanics}, Physical Review Letters,
{\bf74}, 2694--2697, 1995.}
\rif{GC}{1}{\aGC}{0}
 
\def\bGC{ Gallavotti, G., Cohen, E.G.D.: {\it Dynamical ensembles in
stationary states}, in print in Journal of Statistical Physics, 1995.}
\rif{GC}{2}{\bGC}{0}
 
\def\aKr{ Kraichnan, R.H.: {\it Remarks on turbulence}, Advances
in Mathematics, {\bf 16}, 305--331, 1975.}
\rif{Kr}{}{\aKr}{0}
 
\def\aLL{ Landau, L., Lifshitz, V.: {\it M\'ecanique des fluides}, MIR,
Moscow, 1966.}
\rif{LL}{}{\aLL}{0}
 
\def\aMa{ Marchioro, C.: {\it An example of absence of turbulence for
any Reynolds numberb}, Communications in mathematical Physics,
{\bf105}, 99--106, 1986.  And {\it An example of absence of turbulence
for any Reynolds number: $II$}, Communications in mathematical Physics,
{\bf108}, 647--651, 1987.}
\rif{Ma}{}{\aMa}{0}
 
\def\aMP{ Marchioro, C., Presutti, E.: {\it Thermodynamics of particle
systems in the presence of external macroscopic fields: $I$ classical
case}, Communications in Mathematical Physics, {\bf27}, 146-- 154,
1972.}
\rif{MP}{}{\aMP}{0}
 
\def\aMR{ Morriss, G.P., Rondoni, L.: {\it Equivalence of
"nonequilibrium" ensembles for simple maps}, preprint, U.of  N.S. Wales,
Sidney, 1996}
\rif{MR}{}{\aMR}{0}
 
\def\aOt{ Ottino, J.M.: {\it The kinematics of mixing: stretching, chaos
and transport}, Cambridge U. Press, 1989.}
\rif{Ot}{}{\aOt}{0}
 
\def\aPe{ Pedlosky, J.: {\it Geophysical fluid dynamics},
Springer--Verlag, Berlin, 1979.}
\rif{Pe}{}{\aPe}{0}
 
\def\aR{ Ruelle, D.: {\it Chaotic motions and strange attractors},
Lezioni Lincee, notes by S.  Isola, Accademia Nazionale dei Lincei,
Cambridge University Press, 1989; see also: Ruelle, D.: {\it Measures
describing a turbulent flow}, Annals of the New York Academy of
Sciences, {\bf 357}, 1--9, 1980.}
\rif{R}{1}{\aR}{0}
 
\def\bR{ Ruelle, D.: {\it Positivity of entropy production in non
equilibrium statistical mechanics}, Journal of Statistical Physics
{\bf 85}, 1--25, 1996.}
\rif{R}{2}{\bR}{0}

\def\aSm{ Smale, S.: {\it Differentiable dynamical systems}, Bullettin
of the American Mathematical Society, {\bf 73 }, 747--818, 1967.}
\rif{Sm}{}{\aSm}{0}
 
\def\aSJ{ She Z.S., Jackson, E.: {\it Constrained Euler system for
Navier--Stokes turbulence}, Physical Review Letters, {\bf 70},
1255--1258, 1993.}
\rif{SJ}{}{\aSJ}{0}
 
\def\aSEM{ Sarman, S., Evans, D.J., Morriss, G.P.: {\it Conjugate
pairing rule and thermal transport coefficients}, Physical Review,
{\bf42A}, 2233--2242, 1992.}
\rif{SEM}{}{\aSEM}{0}
 
\def\aUF{ Uhlenbeck, G.E., Ford, G.W.: {\it Lectures in Statistical
Mechanics}, American Mathematical society, Providence, R.I.,
pp. 5,16,30, 1963.}
\rif{UF}{}{\aUF}{0}
 
\raf{R}{1}{\aR}{1}
\raf{EM}{}{\aEM}{2}
\raf{ECM}{2}{\bECM}{3}
\raf{BGG}{}{\aBGG}{4}
\raf{Dr}{}{\aDr}{5}
\raf{DM}{}{\aDM}{6}
\raf{SEM}{}{\aSEM}{7}
\raf{BJPV}{}{\aBJPV}{8}
\raf{FP}{}{\aFP}{9}
\raf{GC}{1}{\aGC}{10}
\raf{Ga}{6}{\fGa}{11}
\raf{SJ}{}{\aSJ}{12}
\raf{Pe}{}{\aPe}{13}
\raf{LL}{}{\aLL}{14}
\raf{Ga}{1}{\aGa}{15}
\raf{Kr}{}{\aKr}{16}
\raf{UF}{}{\aUF}{17}
\raf{GC}{2}{\bGC}{18}
\raf{G}{}{\aG}{19}
\raf{Ga}{2}{\bGa}{20}
\raf{MR}{}{\aMR}{21}
\raf{Ga}{3}{\cGa}{22}
\raf{Ga}{4}{\dGa}{23}
\raf{Ga}{5}{\eGa}{24}
\raf{R}{2}{\bR}{25}
\raf{Ge}{}{\aGe}{26}
\raf{Ma}{}{\aMa}{27}
\raf{BG}{}{\aBG}{28}
\raf{Ot}{}{\aOt}{29}
\raf{ECM}{1}{\aECM}{30}
\raf{DPH}{}{\aDPH}{31}
\raf{MP}{}{\aMP}{32}
\raf{ER}{}{\aER}{33}
\raf{Sm}{}{\aSm}{34}
\raf{A}{}{\aA}{35}

\*
\0{\it Internet access:
All the Author's quoted preprints can be found and freely downloaded
(latest postscript version including corrections of misprints and
errors) at:
 
\centerline{\tt http://ipparco.roma1.infn.it}
 
\0in the Mathematical Physics Preprints page.\hfill\break
\sl e-mail address of author: giovanni@ipparco.roma1.infn.it
}

\end